\documentclass[amssymb,aps,prd,floatfix,twocolumn]{revtex4}
\usepackage{amsmath}
\usepackage{amssymb}
\usepackage{graphicx}
\usepackage{tensor}
\usepackage[dvipsnames]{xcolor}

\begin{document}

\title{Accretion of the Vlasov gas on Reissner-Nordstr\"{o}m black holes}

\author{Adam Cie\'{s}lik}
\author{Patryk Mach}\email{patryk.mach@uj.edu.pl}
\affiliation{Instytut Fizyki Teoretycznej, Uniwersytet Jagiello\'{n}ski, \L ojasiewicza 11,  30-348 Krak\'ow, Poland}


\begin{abstract}
We investigate stationary spherically symmetric accretion of the relativistic Vlasov gas on Reissner-Nordstr\"{o}m black holes. The model is based on a recent analysis done by Rioseco and Sarbach for the Schwarzschild spacetime. Both models share many common features: The gas characterized by the Maxwell-J\"{u}ttner distribution at infinity is no longer in thermal equilibrium in the vicinity of the black hole. The radial pressure at the black-hole horizon can be even an order of magnitude smaller than the tangential pressure. Quantitative characteristics of the Reissner-Nordstr\"{o}m model depend on the charge parameter. For black holes with fixed asymptotic mass, the mass accretion rate decreases with the increasing black-hole charge. The ratio of the tangential pressure to the radial pressure at the horizon also decreases with the increasing charge. On the other hand, the particle density at the horizon (normalized by its asymptotic value) grows with the black-hole charge parameter.
\end{abstract}

\maketitle

In this paper we investigate spherical, steady accretion of the relativistic, collisionless (Vlasov) gas on the Reissner-Nordstr\"{o}m black hole. The presented analysis is based on beautiful recent papers by Rioseco and Sarbach \cite{Olivier,Olivier2}, devoted to the accretion of the relativistic Vlasov gas on the Schwarzschild black hole. We decided to generalize the results of Rioseco and Sarbach to Reissner-Nordstr\"{o}m metrics, mostly because of the well-known similarity between causal properties of the Reissner-Nordstr\"{o}m and Kerr spacetimes. This is a common idea, allowing one to retain the simplicity associated with the spherical symmetry and getting an insight into some properties of rotating black holes at the same time (as John Wheeler put it, ``charge is a poor man's angular momentum'').

Theoretical works on accretion date back to early papers by Lyttleton, Hoyle, and Bondi \cite{hoyle_lyttleton, lyttleton_hoyle, bondi_hoyle}, who investigated accretion of dust matter onto a star moving through the interstellar medium. First Newtonian solutions representing spherically symmetric configurations of perfect fluid accretting steadily in the Keplerian gravitational potential were derived by Bondi in \cite{bondi}. This was a crucial conceptual work, defining the physical ingredients of the model---an infinite reservoir of the gas (with fixed nonzero asymptotic density and fixed asymptotic temperature) accreting steadily onto a central object at a rate that is small enough so that the mass of the central object can be treated as constant---and identifying fundamental technical elements---critical solutions, sonic points, etc. Bondi's model was generalized to General Relativity in 1972 by Michel, who considered spherical accretion of the perfect fluid on the Schwarzschild black hole \cite{michel}. Since that time numerous works have been devoted to the analysis of steady, spherical accretion of fluids, assuming different spherically symmetric spacetimes (Reissner-Nordstr\"{o}m, Kottler, Schwarzschild-anti-de Sitter, etc.\ \cite{babichev, mach_malec_karkowski2013, mach2015, ficek}), or taking into account the self-gravity of the fluid \cite{malec1999, karkowski2006, mach_malec2008, dokuchaev}, various equations of state \cite{chaverra2015, chaverra2016}, radiation transfer \cite{radiation1, radiation2}, etc. In all these cases solving the relativistic Euler equation is always an important element of the analysis.

From the technical point of view, the kinetic theory (Vlasov) approach is radically different. On fixed, spherically symmetric spacetimes the Valsov equation can be solved quite generally in terms of suitable canonical coordinates (\cite{Olivier}, but see also \cite{Olivier3,gundlach1,gundlach2,andreasson_rein}). As a consequence, the difficulty of the analysis relies not so much in the actual solving of the Vlasov equation, but rather in finding and analyzing the properties of solutions that would correspond to the gas in thermal equilibrium at infinity, steadily accretting on the black hole. Despite these differences, a comparison with prefect fluids can provide an interpretation of the obtained results. For instance, the eigenvalues of the energy-momentum tensor associated with the Vlasov gas can be interpreted as the energy-density and pressures, but they are no longer degenerate---the pressure does not have to be isotropic.

Repeating the analysis of Rioseco and Sarbach \cite{Olivier} for the Reissner-Nordstr\"{o}m case consisted of two main tasks. On one hand, we had to redo most of the calculations for a general class of spherically symmetric metrics, instead of specifying the analysis to the Schwarzschild solution. On the other hand, the key difficulty was to control the properties of the effective potential for the geodesic motion, which was required to compute the characteristics of the accretion flow expressed in terms of the phase-space integrals. This was still possible for the Reissner-Nordstr\"{o}m metric, although one had to deal with much more complicated formulas. Slightly less complex formulas can be obtained for extremal Reissner-Nordstr\"{o}m spacetimes, so we treat this case separately. In contrast to that, the Kottler or Schwarzschild-de Sitter cases seem to be much more difficult.

The order of this paper is as follows. In Sec.\ \ref{sec:vlasov_spherical_symmetry} we rewrite the formalism derived originally for the Schwarzschild spacetime in \cite{Olivier} for more general, spherically symmetric spacetimes. Since it is a long section, we decided to divide it into several subsections. In Subsection \ref{sec:hamiltonian_description} we recall the Hamiltonian description of the geodesic motion and the Hamiltonian formulation of the Vlasov equation. In Subsection \ref{sec:coordinates} we specify the class of metrics used in the remaining part of the paper and introduce the horizon-penetrating coordinates. A general theory of the Vlasov gas on static, spherically symmetric metrics specified in Subsection \ref{sec:coordinates} is given in Subsection \ref{sec:vlasov_static_spherically_symmetric}. Subsection \ref{sec:maxwell_juttner} provides a short  discussion of the Maxwell-J\"{u}ttner distribution, assumed at infinity. In Subsection \ref{sec:dimensionless} we introduce dimensionless variables and derive the expressions for the observables given in terms of phase-space integrals. In  the last key Subsection \ref{sec:potential} we discuss the properties of the effective potential associated with the Reissner-Nordstr\"{o}m metric and define the regions over which the phase-space integrals are performed. In Sec.\ \ref{sec:gasinequilibrium} we derive analytic expressions for the observables: the particle current density, the energy density, the pressures, etc. Numerical results are collected in Sec.\ \ref{sec:numerical}. Final remarks and conclusions are given in Sec.\ \ref{sec:conclusions}.

Throughout the paper we use geometric units with $c = G = 1$, where $c$ is the speed of light, and $G$ denotes the gravitational constant. The signature of the metric is assumed to be $(-,+,+,+)$. Spacetime dimensions are labeled with Greek indices, $\mu = 0,1,2,3$; spatial dimensions are labeled with Latin indices $i = 1,2,3$.

\section{Vlasov gas in spherical symmetry}
\label{sec:vlasov_spherical_symmetry}

\subsection{Hamiltonian description of the geodesic motion; Vlasov equation}
\label{sec:hamiltonian_description}

\subsubsection{Hamiltonian description}

The relativistic Vlasov gas consists of particles moving along timelike geodesics. Since, following \cite{Olivier}, we use extensively the Hamiltonian formalism, we start by recalling the Hamiltonian description of the geodesic motion. The Hamiltonian of a single particle moving along the geodesic can be chosen as
\[ H = \frac{1}{2} g^{\mu \nu}(x^\alpha) p_\mu p_\nu. \]
Here $(x^\mu, p_\mu)$ are treated as canonical variables, and it is assumed that $H$ depends on $x^\alpha$ through $g^{\mu \nu}(x^\alpha)$. The equations of motion read
\begin{equation}
\label{hamilton_eqs}
\frac{dx^\mu}{d\tau} = \frac{\partial H}{\partial p_\mu}, \quad \frac{dp_\mu}{d\tau} = - \frac{\partial H}{\partial x^\mu}.
\end{equation}
The normalization of the parameter $\tau$ is a matter of convention. We require that $p^\mu = d x^\mu/ d\tau$, and that $H = \frac{1}{2} g^{\mu \nu} p_\mu p_\nu = - \frac{1}{2} m^2$, where $m$ is the particle rest mass. Accordingly, $\tau = s/m$, where $s$ is the proper time. The four velocity $u^\mu = d x^\mu/ ds$ is normalized to minus unity: $g^{\mu \nu}u_\mu u_\nu = -1$.
 
It can be easily shown that Eqs.\ (\ref{hamilton_eqs}) lead to the standard geodesic equation of the form
\[ \frac{d^2 x^\mu}{d \tau^2} + \Gamma^{\mu}_{\alpha \beta} \frac{d x^\alpha}{d\tau} \frac{d x^\beta}{d\tau} = 0, \]
where $\Gamma^\mu_{\alpha \beta}$ denote the Christoffel symbols associated with the metric $g_{\mu \nu}$.

\subsubsection{Vlasov equation}

The relativistic Vlasov equation describes the probability function $f = f(x^\mu,p_\nu)$ \cite{andreasson, rendall}. Since it should be invariant along a geodesic, one requires that
\[ \frac{d}{d\tau} f(x^\mu(\tau),p_\nu(\tau)) = 0, \]
or
\begin{eqnarray*}
\lefteqn{\frac{dx^\mu}{d \tau} \frac{\partial f}{\partial x^\mu} + \frac{d p_\nu}{d \tau} \frac{\partial f}{\partial p_\nu}}\\
&& = \frac{\partial H}{\partial p_\mu} \frac{\partial f}{\partial x^\mu} - \frac{\partial H}{\partial x^\nu} \frac{\partial f}{\partial p_\nu} = \{ H, f \} = 0,
\end{eqnarray*}
where $\{ \cdot, \cdot \}$ denotes the Poisson bracket. The above equation can be written in more explicit terms as
\begin{equation}
\label{vlasov_explicit}
g^{\mu\nu} p_\nu \frac{\partial f}{\partial x^\mu} - \frac{1}{2} p_\alpha p_\beta \frac{\partial g^{\alpha \beta}}{\partial x^\mu} \frac{\partial f}{\partial p_\mu} = 0,
\end{equation}
and it is usually referred to as the relativistic Vlasov equation or the relativistic Liouville equation. For convenience, we chose our phase-space coordinates as $(x^\mu, p_\nu)$, i.e., the coordinates on the cotangent bundle. The version of the Vlasov equation that usually appears in the literature is written in terms of coordinates $(x^\mu, p^\nu)$ (coordinates on the tangent bundle). The transformation from the coordinates $(x^\mu, p_\nu)$ to the new ones $(\tilde x^\mu, \tilde p^\mu)$,
\[ \tilde x^\mu = x^\mu, \quad \tilde p^\mu = g^{\mu \nu}(x^\alpha) p_\nu, \]
yields
\[ \frac{\partial f}{\partial x^\nu} = \frac{\partial f}{\partial \tilde x^\nu} + \frac{\partial g^{\mu \alpha}}{\partial x^\nu} p_\alpha \frac{\partial f}{\partial \tilde p^\mu}, \quad \frac{\partial f}{\partial p_\nu} = g^{\mu \nu} \frac{\partial f}{\partial \tilde p^\mu}. \]
It is easy to show that this change of coordinates transforms Eq.\ (\ref{vlasov_explicit}) to the form
\[ p^\mu \frac{\partial f}{\partial x^\mu} - \Gamma^\mu_{\alpha \beta} p^\alpha p^\beta \frac{\partial f}{\partial p^\mu} = 0, \]
where for simlicity, we have removed the tildes from $(x^\mu, p^\nu)$. An even more common form is obtained by considering a collection of single-mass particles with momenta satisfying the mass shell condition
\begin{equation}
\label{mass_shell}
g_{\mu \nu}p^\mu p^\nu = - m^2.
\end{equation}
In this case, it is sufficient to use the coordinates $(x^\mu, p^i)$, and treat $p^0$ as given by Eq.\ (\ref{mass_shell})---one usually selects the solution corresponding to the future pointing four-momentum. The corresponding transformation from $(x^\mu, p^i)$ to $(\bar x^\mu, \bar p^\mu)$ has the form
\[ \bar x^\mu = x^\mu, \quad \bar p^0 = \bar p^0 (x^\mu, p^i), \quad \bar p^i = p^i, \]
where $p^0 = \bar p^0 (x^\mu, p^i)$ is the solution of Eq.\ (\ref{mass_shell}). This yields
\[ \frac{\partial f}{\partial x^\mu} = \frac{\partial f}{\partial \bar x^\mu} - \frac{1}{p_0} p_\alpha p^\beta \Gamma^\alpha_{\mu \beta} \frac{\partial f}{\partial \bar p^0}, \quad \frac{\partial f}{\partial p^i} = \frac{\partial f}{\partial \bar p^i} - \frac{p_i}{p_0} \frac{\partial f}{\partial \bar p^0}, \]
and after some algebra,
\[ p^\mu \frac{\partial f}{\partial x^\mu} - \Gamma^i_{\alpha \beta} p^\alpha p^\beta \frac{\partial f}{\partial p^i} = 0. \]
Here again, for simplicity, we have removed the bars from $x^\mu, p^i$. Denoting $x^0 = t$, and dividing by $p^0$, we get
\[ \frac{\partial f}{\partial t} + \frac{p^i}{p^0} \frac{\partial f}{\partial x^i} - \frac{1}{p^0} \Gamma^i_{\alpha \beta} p^\alpha p^\beta \frac{\partial f}{\partial p^i} = 0. \]
This is probabaly the most common version of the relativistic Vlassov equation \cite{andreasson, rendall}.

\subsubsection{Integrals over the momentum space}

Many important observable quantities can be expressed as suitable integrals over momenta. The particle current density is given as
\[ J_\mu(x) = \int p_\mu f(x,p) \mathrm{dvol}_x(p), \]
whereas for the components of the energy-momentum tensor one assumes
\[ T_{\mu \nu}(x) = \int p_\mu p_\nu f(x,p) \mathrm{dvol}_x(p), \]
where the momentum-space integration element is given by 
\begin{eqnarray}
\mathrm{dvol}_x(p) & = & \sqrt{- \det [g^{\mu \nu}(x)]} d^4 p \nonumber \\
& = & \sqrt{- \det [g^{\mu \nu}(x)]} dp_0 dp_1 dp_2 dp_3.
\label{momentumvolume}
\end{eqnarray}

Using Eq.\ (\ref{vlasov_explicit}) one can easily show that the particle current density $J^\mu$ satisfies the conservation equation \cite{rezzolla_zanotti}
\begin{equation}
\label{conservationJ}
\nabla_\mu J^\mu = 0.
\end{equation}

\subsection{Horizon-penetrating coordinates}
\label{sec:coordinates}

Although in this paper we work ultimately with the Reissner-Nordstr\"{o}m spacetime, many formulas derived in the following sections hold for general spherically symmetric metrics of the form
\begin{eqnarray}
g  & = & g_{tt}(r) dt^2 + 2 g_{tr}(r) dt dr + g_{rr}(r) dr^2 \nonumber \\
&& + r^2 (d \theta^2 + \sin^2 \theta d\varphi^2).
\label{metric_standard}
\end{eqnarray}
On the other hand, for computational convenience, the majority of the formulas will be obtained for the spherically symmetric metrics which in some coordinate system can be written as
\begin{equation}
g = - N(\bar r) d\bar t^2 + \frac{1}{N(\bar r)} d \bar r^2 + \bar r^2 (d \theta^2 + \sin^2 \theta d \varphi^2)
\label{metric_diagonal}
\end{equation}
(Schwarzschild, Reissner-Nordstr\"{o}m, Kottler, and Schwarzschild-anti-de Sitter metrics belong to this category). Since the coordinate system used in (\ref{metric_diagonal}) is divergent at the black-hole horizon, we will instead work in horizon-penetrating, Eddington-Finkelstein type coordinates. For a metric of the form (\ref{metric_diagonal}) we define a new time coordinate $t = t(\bar t, \bar r)$,
\[ t = \bar t + \int^{\bar r} \left[ \frac{1}{N(r)} - \eta(r) \right] dr,  \] 
keeping the areal radius as a new radial coordinate $r = \bar r$. Here $\eta(r)$ is an arbitrary function. This yields
\begin{eqnarray}
g & = & -N dt^2 + 2(1 - N \eta) dt dr + \eta (2 - N \eta) dr^2 \nonumber \\
&& + r^2 (d \theta^2 + \sin^2 \theta d \varphi^2).
\label{efgeneral}
\end{eqnarray}
The corresponding contravariant metric components read
\[ g^{tt} = \eta (-2 + N \eta), \quad g^{tr} = 1 - N \eta, \quad g^{rr} = N. \]
Note that
\begin{equation}
\label{metricunity}
\left(g^{tr}\right)^2 - g^{rr}g^{tt} = 1.
\end{equation}

In particular, the Reissner-Nordstr\"{o}m metric can be written as
\begin{eqnarray*}
g & = & - \left( 1 - \frac{2M}{\bar r} + \frac{Q^2}{\bar r^2} \right) d \bar t + \left( 1 - \frac{2M}{\bar r} + \frac{Q^2}{\bar r^2} \right)^{-1} d \bar r^2 \\
& & + \bar r^2 (d \theta^2 + \sin^2 \theta d \varphi^2).
\end{eqnarray*}
Taking $\eta \equiv 1$ (this choice is sometimes referred to as Eddington-Finkelstein coordinates), we obtain
\begin{eqnarray}
g & = & - \left( 1 - \frac{2M}{r} + \frac{Q^2}{r^2} \right) dt^2 + 2 \left( \frac{2M}{r} - \frac{Q^2}{r^2} \right) dt dr \nonumber \\
&& + \left( 1 + \frac{2M}{r} - \frac{Q^2}{r^2} \right) dr^2 + r^2 (d \theta^2 + \sin^2 \theta d \varphi^2).
\label{rn_eddington_finkelstein}
\end{eqnarray}
The contravariant metric components corresponding to metric (\ref{rn_eddington_finkelstein}) read
\begin{subequations}
\label{rncontravariant}
\begin{eqnarray}
g^{tt} & = & -1 - \frac{2M}{r} + \frac{Q^2}{r^2}, \\ g^{tr} & = & \frac{2M}{r} - \frac{Q^2}{r^2}, \\
g^{rr} & = & 1 - \frac{2M}{r} + \frac{Q^2}{r^2}.
\end{eqnarray}
\end{subequations}

In the following, we will distinguish the formulas valid for the metric (\ref{metric_standard}) from those obtained for (\ref{efgeneral}) by a suitable choice of the notation. In the former case, we will use the metric components $g_{tt}$, $g_{tr}$, $g_{rr}$ (or their contravariant counterparts). In the latter, we will write the formulas in terms of the functions $N$ and $\eta$. Although the components of the vector and tensor quantities depend explicitly on the choice of the time foliation (the choice of $\eta$ in our case), important physical quantities (the particle density, the energy density, the pressures, etc.) are independent of $\eta$. We will try to emphasise this fact by writing the corresponding formulas in a way manifestly independent of $\eta$.

\subsection{Vlasov equation on static spherically symmetric spacetimes}
\label{sec:vlasov_static_spherically_symmetric}

\subsubsection{Conserved quantities}

For the general spherically symmetric metric of the form (\ref{metric_standard}) the Hamiltonian of a free particle can be written as
\begin{eqnarray}
H & = & \frac{1}{2} \left[ g^{tt}(r) (p_t)^2 + 2 g^{tr}(r) p_t p_r + g^{rr}(r) (p_r)^2 \right. \nonumber \\
&&  \left. + \frac{1}{r^2} (p_\theta)^2 + \frac{1}{r^2 \sin^2 \theta} (p_\varphi)^2 \right].
\label{hamiltonian1}
\end{eqnarray}
Since $H$ does not depend on $t$ and $\varphi$, the momenta  $E \equiv -p_t$, $l_z \equiv p_\varphi$ are constants of motion. Moreover, as the Hamiltonian $H$ does not depend explicitly on $\tau$, it is a constant of motion itself. Because $H = -\frac{1}{2}m^2$, this is another expression of the fact that the rest mass of a free particle is constant. Less obvious is that
\begin{equation}
\label{constraintl}
l = \sqrt{p_\theta^2 + \frac{p_\varphi^2}{\sin^2 \theta}}
\end{equation}
is also a constant of motion---this fact follows directly from the assumed spherical symmetry. In terms of the above constants, Eq.\ (\ref{hamiltonian1}) can be written as
\begin{equation}
\label{constraintm}
 g^{tt}(r) E^2 - 2 g^{tr}(r) E p_r + g^{rr}(r) (p_r)^2 + \frac{l^2}{r^2} + m^2 = 0.
\end{equation}

\subsubsection{Classification of trajectories}

Solving Eq.\ (\ref{constraintm}) with respect to $p_r$ one gets
\begin{eqnarray}
p_r & = & \frac{g^{tr} E \pm \sqrt{\left[ (g^{tr})^2 - g^{tt} g^{rr} \right] E^2 - g^{rr} \left( m^2 + \frac{l^2}{r^2} \right)}}{g^{rr}} \nonumber \\
& = & \frac{(1 - N \eta) E \pm \sqrt{E^2 - \tilde U_{l,m}(r)}}{N},
\label{prformula}
\end{eqnarray}
where we have used Eq.\ (\ref{metricunity}) and introduced the effective potential
\[ \tilde U_{l,m}(r) = N \left( m^2 + \frac{l^2}{r^2} \right). \]
For the Reissner-Nordstr\"{o}m metric we have $\tilde U_{l,m}(r) \to m^2$, as $r \to \infty$. Consequently, only trajectories with $E \ge m$ can reach infinity.

In the spherical accretion problem we are naturally interested in particle trajectories that originate at infinity and go inward, attracted by the black hole. This family of trajectories can be further divided into two subclasses, crucial in the following analysis: those absorbed by the black hole, denoted with (abs), and those scattered to the infinity, marked with (scat). The division into those subclasses depends on the properties of the effective potential $\tilde U_{l,m}(r)$. A trajectory originating at infinity with the angular momentum sufficiently high, so that at some finite distance $E^2 - \tilde U_{l,m}(r) = 0$, is reflected backward to infinity. Otherwise, it can reach the black hole. A precise characterization of both classes of absorbed and scattered trajectories is important and will be studied for the Reissner-Nordstr\"{o}m spacetimes in Sec.\ \ref{sec:potential}.

\subsubsection{Action-angle variables}

A convenient way of proceeding further is to introduce suitably defined action-angle variables \cite{Olivier}. This is a standard procedure in classical mechanics \cite{Hand,Goldstein}, however the details of the transformation used here (and in \cite{Olivier}) are subtle.

Let $\gamma$ be a geodesic orbit with constant $m$, $E$, $l_z$, and $l$, joining some reference point with a point with coordinates $(t,r,\theta,\varphi)$. We introduce a generating function (the abbreviated action)
\begin{equation}
\label{actionS}
S = - E t + l_z \varphi + \int_\gamma p_r dr + \int_\gamma p_\theta d \theta,
\end{equation}
where the integrals are understood as line integrals along the orbit $\gamma$.  More precisely, the first integral in Eq.\ (\ref{actionS}) is the line integral along the projection of the orbit $\gamma$ onto the $(r,p_r)$ plane; the second integral is performed along the projection of $\gamma$ onto the $(\theta, p_\theta)$ plane. Thus, $p_\theta$ in Eq.\ (\ref{actionS}) can be expressed as
\begin{equation}
\label{ptheta}
p_\theta = \pm \sqrt{l^2 - \frac{l_z^2}{\sin^2 \theta}},
\end{equation}
while $p_r$ is given by Eq.\ (\ref{prformula}). Note that the integral
\[ \int_\gamma p_\theta d\theta = \pm \int \sqrt{l^2 - \frac{l_z^2}{\sin^2 \theta}} d\theta, \]
can be actually computed analytically. Possible choices of the starting (or reference) points of orbits $\gamma$ are discussed in \cite{Olivier}.

We define a canonical transformation taking as new momenta, the constants
\begin{widetext}
\begin{eqnarray*}
P_0 & = & m = \sqrt{- g^{tt}(r) (p_t)^2 - 2 g^{tr}(r) p_t p_r - g^{rr}(r) (p_r)^2 - \frac{1}{r^2} \left( p_\theta^2 + \frac{p_\varphi^2}{\sin^2 \theta} \right)}, \\
P_1 & = & E = - p_t, \\
P_2 & = & l_z = p_\varphi, \\
P_3 & = & l = \sqrt{p_\theta^2 + \frac{p_\varphi^2}{\sin^2 \theta}}. 
\end{eqnarray*}
\end{widetext}
Then the corresponding conjugate variables are defined as
\begin{subequations}
\begin{eqnarray}
\label{q0}
Q^0 = \frac{\partial S}{\partial m} & = & - m \int_\gamma \frac{dr}{-g^{tr} E + g^{rr} p_r}, \\
\label{q1}
Q^1 = \frac{\partial S}{\partial E} & = & - t + \int_\gamma \frac{g^{tt} E - g^{tr} p_r}{g^{tr} E - g^{rr} p_r} dr, \\
\label{q2}
Q^2 = \frac{\partial S}{\partial l_z} & = & \varphi - l_z \int_\gamma \frac{d \theta}{p_\theta \sin^2 \theta}, \\
Q^3 = \frac{\partial S}{\partial l} & = & - l \int_\gamma \frac{dr}{r^2 \left( -g^{tr} E + g^{rr} p_r \right)} \nonumber \\
&& + l \int_\gamma \frac{d \theta}{p_\theta}.
\label{q3}
\end{eqnarray}
\end{subequations}
Here again, all integrals are  understood as line integrals along trajectories with fixed $m, E, l_z, l$. It is important to keep this in mind when considering the transformation $(t,r,\theta,\phi,p_t,p_r,p_\theta,p_\varphi) \to (Q^\mu,P_\nu)$ as a coorindate transformation in the phase space.

In terms of the action-angle variables $(P_\mu, Q^\nu)$, the Hamiltonian reads simply $H = -P_0^2/2$. Since the Poisson bracket is covariant with respect to canonical transformations, we have
\begin{eqnarray*}
\lefteqn{\frac{\partial H}{\partial p_\mu} \frac{\partial}{\partial x^\mu} - \frac{\partial H}{\partial x^\nu} \frac{\partial}{\partial p_\nu}} \\
&& = \frac{\partial H}{\partial P_\mu} \frac{\partial}{\partial Q^\mu} - \frac{\partial H}{\partial Q^\nu} \frac{\partial}{\partial P_\nu} = - P_0 \frac{\partial}{\partial Q^0}.
\end{eqnarray*}
Accordingly, the Vlasov equation takes the form
\[ \frac{\partial f}{\partial Q^0} = 0, \]
and its general solution can be written as
\begin{equation}
\label{general_solution}
f(x^\mu, p_\nu) = \mathcal F (Q^1, Q^2, Q^3, P_0, P_1, P_2, P_3).
\end{equation}
Further restrictions on the distribution function $f$ can be given assuming symmetry conditions.

\subsubsection{Symmetries}

In the following we impose the conditions of stationarity, and spherical symmetry. Here the key step is to compute the lifts of the Killing vectors generating the symmetries to the cotangent bundle with the local coordinates $(x^\mu, p_\nu)$ or $(Q^\mu, P_\nu)$. For the Killing vector
\[ \xi_x = \left. \xi^\mu (x) \frac{\partial}{\partial x^\mu} \right|_x \]
we compute the lifts as
\[ \hat \xi_{(x,p)} = \left. \xi^\mu (x) \frac{\partial}{\partial x^\mu}\right|_{(x,p)} - \left. p_\alpha \frac{\partial \xi^\alpha}{\partial x^\mu}(x) \frac{\partial}{\partial p_\mu} \right|_{(x,p)}.  \]

The Killing vectors generating the action of the rotation group SO(3) on the spacetime can be given as
\[ \xi_1 + i \xi_2 = e^{i \varphi} \left( i \frac{\partial}{\partial \theta} - \cot \theta \frac{\partial}{\partial \varphi} \right), \quad \xi_3 = \frac{\partial}{\partial \varphi}, \]
where a convenient complex-number notation is used to simplify the formulas. The Killing vector generating time-translations is simply
\[ k = \frac{\partial}{\partial t}. \]
It is easily to check that the lifts of these Killing vectors to the cotangent bundle are given by
\[ \hat \xi_1 + i \hat \xi_2 = e^{i \varphi} (\hat \eta_1 + i \hat \eta_2), \quad \hat \xi_3 = \frac{\partial}{\partial \varphi}, \quad \hat k = \frac{\partial}{\partial t}, \]
where
\begin{eqnarray*}
\hat \eta_1 + i \hat \eta_2 & = & i \frac{\partial}{\partial \theta} - \cot \theta \frac{\partial}{\partial \varphi} - \frac{p_\varphi}{\sin^2 \theta} \frac{\partial}{\partial p_\theta} \\
&& + (p_\theta + i p_\varphi \cot \theta) \frac{\partial}{\partial p_\varphi}.
\end{eqnarray*}

Since the general solution of the Vlasov equation (\ref{general_solution}) is given conveniently in terms of coordinates $(Q^\mu, P_\nu)$, we express $\hat k$, $\hat \xi_1$, $\hat \xi_2$, $\hat \xi_3$ also in terms of $(Q^\mu, P_\nu)$. The formulas for $\hat k$ and $\hat \xi_3$ are simple:
\[ \hat \xi_3 = \frac{\partial}{\partial \varphi} = \frac{\partial Q^\mu}{\partial \varphi} \frac{\partial}{\partial Q^\mu} + \frac{\partial P_\nu}{\partial \varphi} \frac{\partial}{\partial P_\nu} = \frac{\partial}{\partial Q^2}, \]
\[ \hat k = \frac{\partial}{\partial t} = \frac{\partial Q^\mu}{\partial t} \frac{\partial}{\partial Q^\mu} + \frac{\partial P_\nu}{\partial t} \frac{\partial}{\partial P_\nu} = - \frac{\partial}{\partial Q^1.} \]
The formulas for $\hat \eta_1$ and $\hat \eta_2$ are more complex:
\begin{widetext}
\begin{eqnarray}
\hat \eta_1 + i \hat \eta_2 & = & (p_\theta + i l_z \cot \theta) \frac{\partial}{\partial P_2} \nonumber \\
&& + \left[ -\cot \theta - p_\theta l^2 \int_\gamma \frac{d \theta}{p_\theta^3 \sin^2 \theta} + i \left( - \frac{l_z}{p_\theta \sin^2 \theta} - l_z l^2 \cot \theta \int_\gamma \frac{d \theta}{p^3_\theta \sin^2 \theta} \right) \right] \frac{\partial}{\partial Q^2} \nonumber \\
&& + \left[ p_\theta l_z l \int_\gamma \frac{d\theta}{p^3_\theta \sin^2 \theta} + i \left( \frac{l}{p_\theta} + l_z^2 l \cot \theta \int_\gamma \frac{d \theta}{p_\theta^3 \sin^2 \theta} \right) \right] \frac{\partial}{\partial Q^3}.
\end{eqnarray}
\end{widetext}
The above result is derived as follows. First we note that both $\eta_1$ and $\eta_2$ leave the constants $m$, $E$ and $l$ (but not $l_z$) invariant, i.e., $(\eta_1 + i \eta_2)(m) = (\eta_1 + i \eta_2)(E) = (\eta_1 + i \eta_2)(l) = 0$. Also, since $Q^1$ does not depend on $\theta, \varphi, p_\theta,$ or $p_\varphi$, and $Q^0$ depends on $\theta$, $p_\theta$, and $p_\varphi$ only through $m$, we also have $(\eta_1 + i \eta_2)(Q^0) = (\eta_1 + i \eta_2)(Q^1) = 0$. Thus, $\eta_1 + i \eta_2$ are only spanned by $\partial/\partial P_2$, $\partial/\partial Q^2$, $\partial/\partial Q^3$. Computing $(\eta_1 + i \eta_2)(Q^2)$ and $(\eta_1 + i \eta_2)(Q^3)$ is then straightforward, keeping in mind that $p_\theta$ under integral signs in Eqs.\ (\ref{q2},\ref{q3}) is given by  Eq.\ (\ref{ptheta}).

We see that a distribution function $f$ with a connected support in the phase space is stationary, if it is independent of $Q^1$, axially symmetric, if it is independent of $Q^2$, and spherically symmetric, if it is independent of $P_2$, $Q^2$, and $Q^3$. Note that the above statement is not generally (i.e, without additional assumptions on $f$) true. A counterexample can be found in \cite{schaeffer}.

In what follows, we restrict ourselves to stationary and spherically symmetric solutions of the form
\begin{equation}
\label{stationary_sph_sym_solution}
f(x^\mu, p_\nu) = \mathcal F (P_0, P_1, P_3).
\end{equation}

\subsection{The gas in thermal equilibrium at infinity}
\label{sec:maxwell_juttner}

At this stage, specifying a solution corresponding to the gas in thermal equilibrium at infinity is quite simple. In the flat spacetime the distribution function $f$ describing the relativistic, nondegenerate gas in thermal equilibrium is known as the J\"{u}ttner or Maxwell-J\"{u}ttner distribution, and it is a relativistic counterpart of the Maxwell distribution \cite{juttner1, juttner2, israel}. For the so-called simple gas (the gas of same mass particles) it can be written as
\begin{equation}
\label{juttner_original}
f(x^\mu,p_\nu) = \alpha \delta \left( \sqrt{- p^\mu p_\mu} - m \right) e^{-\beta E},
\end{equation}
where $m$ is the particle mass, and $E = -p_t$ is the particle energy. Here $\alpha$ is a normalization constant, and $\beta = (k_\mathrm{B} T)^{-1}$, where $T$ is the temperature and $k_\mathrm{B}$ denotes the Boltzmann constant. The normalization constant $\alpha$ can be related with the particle density given by
\[ n_\infty(z) = 4 \pi \alpha m^4 \frac{K_2(z)}{z}, \]
where $z = m/(k_\mathrm{B} T)$, and $K_2$ is the modiefied Bessel functions of the second kind \cite{israel}.

Returning to the spherically symmetric, asymptotically flat metrics of the form (\ref{metric_standard}), we write the distribution (\ref{juttner_original}) in terms of the coordinates $(Q^\mu,P_\nu)$ as
\begin{equation}
\label{juttnerPQ}
f = \alpha \delta(P_0 - m) e^{-\beta P_1}.
\end{equation}
The above formula constitutes a spherically symmetric stationary solution of the Vlasov equation valid everywhere, not only at the infinity. However, it describes a gas in thermal equilibrium only asymptotically (i.e., for the flat metric). We should emphasise that for a finite radius, the parameter $T$ can no longer be associated with the temperature.

In what follows, we will generally assume distributions of the form (\ref{stationary_sph_sym_solution}) and later specialize to (\ref{juttnerPQ}). Thus, the whole difficulty in describing the spherically symmetric accretion of the Vlasov gas is effectively reduced to the computation of relevant (observable) quantities.

\subsection{Dimensionless variables; phase-space integrals}
\label{sec:dimensionless}

\subsubsection{Dimensionless variables}

Following \cite{Olivier} we introduce dimensionless variables: $\tau$, $\xi$, $\pi_\xi$, $\pi_\theta$, $\varepsilon$, $\lambda$, $\lambda_z$ as follows
\begin{eqnarray*}
t & = & M \tau, \\
r & = & M \xi, \\
p_r & = & m \pi_\xi, \\
p_\theta & = & M m \pi_\theta, \\
E & = & m \varepsilon, \\
l & = & M m \lambda, \\
l_z & = & M m \lambda_z. 
\end{eqnarray*}
In addition, we define the charge parameter $q$ as $q = Q/M$. Metric components (\ref{rncontravariant}) can be expressed in terms of the dimensionless variables as
\begin{eqnarray*}
g^{tt} & = & - \left( 1 + \frac{2}{\xi} - \frac{q^2}{\xi^2} \right), \\
g^{tr} & = & \frac{2}{\xi} - \frac{q^2}{\xi^2}, \\
g^{rr} & = & 1 - \frac{2}{\xi} + \frac{q^2}{\xi^2}.
\end{eqnarray*}
The two horizons of the Reissner-Nordstr\"{o}m spacetime are located at $\xi_\pm = 1 \pm \sqrt{1 - q^2}$. For the variables $Q^\mu$ we get
\begin{eqnarray*}
Q^0 & = & - M \int_\gamma \frac{d \xi}{-g^{tr} \varepsilon + g^{rr} \pi_\xi}, \\
Q^1 & = & M \left( - \tau + \int_\gamma \frac{g^{tt} \varepsilon - g^{tr} \pi_\xi}{g^{tr} \varepsilon - g^{rr} \pi_\xi} d\xi \right), \\
Q^2 & = & \varphi - \lambda_z \int_\gamma \frac{d \theta}{\pi_\theta \sin^2 \theta}, \\
Q^3 & = & - \lambda \int_\gamma \frac{d \xi}{\xi^2 \left( - g^{tr} \varepsilon + g^{rr} \pi_\xi \right)} + \lambda \int_\gamma \frac{d \theta}{\pi_\theta}.
\end{eqnarray*}
The two constraint equations (\ref{constraintm}) and (\ref{constraintl}) read, respectively,
\[ g^{tt} \varepsilon^2 - 2 g^{tr} \varepsilon \pi_\xi + g^{rr} \pi_\xi^2 + \frac{\lambda^2}{\xi^2} + 1 = 0 \]
and
\begin{equation}
\label{constraintlambda}
\lambda^2 = \pi_\theta^2 + \frac{\lambda_z^2}{\sin^2 \theta}.
\end{equation}
Consequently for the radial momentum we get
\[ \pi_{\xi \pm} = \frac{(1 - N \eta) \varepsilon \pm \sqrt{\varepsilon^2 - U_\lambda(\xi)}}{N}, \]
where the dimensionless effective potential $U_\lambda(\xi)$ is given as
\[ U_\lambda(\xi) = N\left( 1 + \frac{\lambda^2}{\xi^2} \right). \]
In what follows, we will also denote
\[ \epsilon(\pi_{\xi \pm}) = \pm 1. \]

Note that these formulas are general, valid for any metric of the form (\ref{efgeneral}). Also note, that the form of the effective potential $U_\lambda(\xi)$ does not depend on the particular gauge, which in our case is specified by the choice of the function $\eta$. For the Reissner-Nordstr\"{o}m metric we have
\[  U_\lambda(\xi) = \left( 1 - \frac{2}{\xi} + \frac{q^2}{\xi^2} \right) \left( 1 + \frac{\lambda^2}{\xi^2} \right). \]

Since the expression $\sqrt{\varepsilon^2 - U_\lambda(\xi)}$ will appear frequently in the remaining part of this paper, we will denote it with
\[ s_\lambda(\varepsilon,\xi) = \sqrt{\varepsilon^2 -  U_\lambda(\xi)}. \]
Note that
\[ s_0(\varepsilon, \xi) = \sqrt{\varepsilon^2 - N}. \]
Also for $N \to 0$ (at the horizon), we have $s_\lambda \to \varepsilon$.

\subsubsection{Momentum-space volume element}

In the following, we introduce another variant of momentum coordinates, suitable for expressing the integration element (\ref{momentumvolume}). It is convenient to choose the set $(\varepsilon, m, \lambda, \chi)$, where $\chi$ is chosen as a momentum coordinate compatible with the constraint equation (\ref{constraintlambda}). We define
\[ \pi_\theta = \lambda \cos \chi, \quad \lambda_z = \lambda \sin \theta \sin \chi,  \]
and change the variables $(\pi_\theta, \lambda_z)$ to $(\lambda, \chi)$. In total, we change the momentum variables from $(p_t,p_r,p_\theta,p_\varphi)$ to $(\varepsilon, m, \lambda, \chi)$, according to
\[ p_t = -m \varepsilon, \quad p_\theta = M m \lambda \cos \chi, \quad p_\varphi = M m \lambda \sin \theta \sin \chi. \]
The radial momentum $p_r$ is given as a solution to the equation
\[ g^{tt} m^2 \varepsilon^2 - 2 g^{tr} m \varepsilon p_r + g^{rr} (p_r)^2 + \frac{m^2 \lambda^2}{\xi^2} + m^2 = 0. \]
Computing
\[ \frac{\partial(p_t,p_r,p_\theta,p_\varphi)}{\partial(\varepsilon, m, \lambda, \chi)} = \frac{M^2 m^3 \lambda \sin \theta}{g^{rr} \pi_\xi - g^{tr} \varepsilon} = \epsilon(\pi_\xi) \frac{M^2 m^3 \lambda \sin \theta}{\sqrt{\varepsilon^2 - U_\lambda(\xi)}} \]
and
\[ \sqrt{- \det [g^{\mu \nu} (x)]} = \frac{1}{r^2 \sin \theta}, \]
we get
\[ \mathrm{dvol}_x(p) = \frac{1}{\xi^2} \frac{m^3 \lambda}{ \sqrt{\varepsilon^2 - U_\lambda (\xi)}} d \varepsilon dm d\lambda d \chi. \]
Note that the above compact form assumes the metric of the form (\ref{efgeneral}).

\subsubsection{Momentum integrals}
\label{momentumintegrals}

We consider static, sphericall symmetric distributions with
\[ f(x^\mu, p_\nu) = \mathcal F (P_0, P_1, P_3) = \mathcal F (m, m \varepsilon, M m \lambda). \]
Following \cite{Olivier}, let us introduce the following abbreviation allowing to perform the integration over $m$:
\begin{equation}
\label{massmoment}
\mathcal F_n (\varepsilon, \lambda) = \int_0^\infty m^n \mathcal F (m, m \varepsilon, M m \lambda) d m.
\end{equation}
We begin by computing the particle current density
\[ J_\mu(\xi) = \frac{1}{\xi^2} \int \frac{p_\mu \mathcal F(m, m\varepsilon, M m \lambda) m^3 \lambda}{\sqrt{\varepsilon^2 - U_\lambda (\xi) }} d\varepsilon dm d\lambda d\chi.  \]
Since $p_\theta = M m \lambda \cos \chi$ and $p_\varphi = M m \sin \theta \lambda \sin \chi$, we get $J_\theta = 0$ and $J_\varphi = 0$. This follows immediately, by evaluating the integrals with respect to $\chi$ over the entire period $(0,2\pi)$. For the two non-zero components $J_t$ and $J_r$ we get
\begin{eqnarray*}
J_t (\xi) & = &  \frac{2 \pi}{\xi^2} \int \frac{p_t \mathcal F(m, m\varepsilon, M m \lambda) m^3 \lambda}{\sqrt{\varepsilon^2 - U_\lambda (\xi) }} d\varepsilon dm d\lambda \\
& = &  - \frac{2 \pi}{\xi^2} \int \frac{\varepsilon \mathcal F_4(\varepsilon, \lambda) \lambda}{\sqrt{\varepsilon^2 - U_\lambda (\xi) }} d\varepsilon d\lambda
\end{eqnarray*}
and
\begin{eqnarray*}
J_r (\xi) & = &  \frac{2 \pi}{\xi^2} \int \frac{p_r \mathcal F(m, m\varepsilon, M m \lambda) m^3 \lambda}{\sqrt{\varepsilon^2 - U_\lambda (\xi) }} d\varepsilon dm d\lambda \\
& = & \frac{2 \pi}{\xi^2} \int \frac{\pi_\xi \mathcal F_4(\varepsilon, \lambda) \lambda}{\sqrt{\varepsilon^2 - U_\lambda (\xi) }} d\varepsilon d\lambda.
\end{eqnarray*}
A simple computation shows that
\[ J^r = g^{rt} J_t + g^{rr} J_r = \frac{2 \pi}{\xi^2} \int \epsilon (\pi_\xi) \mathcal F_4 (\varepsilon, \lambda) \lambda d \varepsilon d \lambda. \]
This means that $4 \pi r^2 J^r = \mathrm{const}$, i.e., the flux of particles through a sphere of radius $r$ is independent of $r$. Of course, it is also a direct consequence of the conservation law (\ref{conservationJ}). The same result is true also for the standard hydrodynamic accretion. We define the (baryonic) mass accretion rate as
\begin{equation}
\label{mdotdef}
\dot M \equiv - 4 \pi m r^2 J^r.
\end{equation}

The energy momentum tensor can be computed as
\[ T_{\mu \nu}(\xi) = \frac{1}{\xi^2} \int \frac{p_\mu p_\nu \mathcal F(m, m\varepsilon, M m \lambda) m^3 \lambda}{\sqrt{\varepsilon^2 - U_\lambda (\xi) }} d\varepsilon dm d\lambda d\chi.  \]
By evaluating the integrals with respect to $\chi$, one can show that $T_{\theta \varphi} = 0$, $T_{t \theta} = 0$, $T_{r \theta} = 0$, $T_{t \varphi} = 0$, $T_{r \varphi} = 0$. Similarly, we can show that $T_{\varphi \varphi} = \sin^2 \theta T_{\theta \theta},$
where
\begin{eqnarray}
\nonumber
T_{\theta \theta}(\xi) & = & \frac{\pi M^2}{\xi^2} \int \frac{\mathcal F(m, m\varepsilon, M m \lambda) m^5 \lambda^3}{\sqrt{\varepsilon^2 - U_\lambda(\xi)}} d\varepsilon dm d\lambda \\
& = & \frac{\pi M^2}{\xi^2} \int \frac{\mathcal F_5(\varepsilon, \lambda) \lambda^3}{\sqrt{\varepsilon^2 - U_\lambda(\xi)}} d\varepsilon d\lambda.
\label{Tthetatheta}
\end{eqnarray}
In the same way, we express
\begin{eqnarray}
\label{Ttt}
T_{tt} & = & \frac{2 \pi}{\xi^2} \int \frac{\mathcal F_5(\varepsilon, \lambda) \varepsilon^2 \lambda}{\sqrt{\varepsilon^2 - U_\lambda(\xi)}} d\varepsilon d \lambda, \\
\label{Ttr}
T_{tr} & = & - \frac{2 \pi}{\xi^2} \int \frac{\pi_\xi \mathcal F_5(\varepsilon, \lambda) \varepsilon \lambda}{\sqrt{\varepsilon^2 - U_\lambda(\xi)}} d\varepsilon d \lambda, \\
\label{Trr}
T_{rr} & = & \frac{2 \pi}{\xi^2} \int \frac{\pi_\xi^2 \mathcal F_5(\varepsilon, \lambda) \lambda}{\sqrt{\varepsilon^2 - U_\lambda(\xi)}} d\varepsilon d \lambda.
\end{eqnarray}

In order to proceed further, one has to specify the metric. There are two technical reasons for that. Clearly, one has to specify the effective potential $U_\lambda(\xi)$, which appears in the above integrals (both explicitly and in the expression for $\pi_\xi$). More importantly, the knowledge of $U_\lambda(\xi)$ is essential to establish the regions in the momentum space over which the above integrations are performed. This requires a subtle analysis, which we do in the next section.

\subsection{Properties of the effective potential}
\label{sec:potential}

\subsubsection{Extremal Reissner-Nordstr\"{o}m metric}

For clarity, we start with the formulas valid for the extremal Reissner-Nordstr\"{o}m metric. In this case
\[ U_\lambda(\xi) = \left( 1 - \frac{2}{\xi} + \frac{1}{\xi^2} \right) \left( 1 + \frac{\lambda^2}{\xi^2} \right) = \frac{(\xi - 1)^2}{\xi^2} \left( 1 + \frac{\lambda^2}{\xi^2} \right). \]
The derivative $dU_\lambda/d\xi$ reads
\[ \frac{dU_\lambda}{d\xi} = \frac{2(\xi - 1)(\xi^2 + 2 \lambda^2 - \lambda^2 \xi)}{\xi^5}. \]
There is a local minimum of $U_\lambda(\xi)$ at $\xi = 1$, i.e., at the horizon and, if $\lambda^2 > 8$, a maximum at
\begin{equation}
\label{ximax}
 \xi_\mathrm{max} = \frac{\lambda^2}{2} \left(1 - \sqrt{1 - \frac{8}{\lambda^2}} \right)
 \end{equation}
and a local minimum at
\[ \xi_\mathrm{min} = \frac{\lambda^2}{2} \left(1 + \sqrt{1 - \frac{8}{\lambda^2}} \right). \]
For $\xi \to \infty$ the potential $U_\lambda(\xi) \to 1$. It is easy to check that for $\lambda^2$ growing from $\lambda^2 = 8$ to infinity, the location of the local maximum $\xi_\mathrm{max}$ decreases from $\xi_\mathrm{max} = 4$ to $\xi_\mathrm{max} = 2$. On the other hand, $\xi_\mathrm{min}$ grows from $\xi_\mathrm{min} = 4$ for $\lambda^2 = 8$ to infinity for $\lambda^2 \to \infty$.

At $\xi_\mathrm{min}$ and $\xi_\mathrm{max}$ we have $\xi^2 = \lambda^2 (\xi - 2)$. Consequently, at $\xi_\mathrm{min}$ or $\xi_\mathrm{max}$,
\begin{equation}
\label{u_at_max}
  U_\lambda(\xi)  = \frac{(\xi - 1)^2}{\xi^2} \left( 1 + \frac{1}{\xi - 2} \right) = \frac{(\xi - 1)^3}{\xi^2 (\xi - 2)}.
  \end{equation}
It follows that $U_\lambda(\xi_\mathrm{max})$ grows from 27/32 for $\lambda^2 = 8$ to infinity for $\lambda^2 \to \infty$. At the same time $U_\lambda(\xi_\mathrm{min})$ grows from 27/32 for $\lambda^2 = 8$ to 1 for $\lambda^2 \to \infty$. It is also important to note that $U_\lambda(\xi_\mathrm{max}) = 1$ for $\xi_\mathrm{max} = (3 + \sqrt{5})/2 \approx 2.61803$ (this follows immediately from Eq.\ (\ref{u_at_max})).

The next important quantity is the value of the angular momentum $\lambda$ for which the value $U_\lambda(\xi_\mathrm{max})$ is equal to a given value $\varepsilon^2 > 1$. It can be computed by solving the cubic equation
\begin{equation}
\label{forlambdac}
\frac{(\xi_\mathrm{max} - 1)^3}{\xi_\mathrm{max}^2 (\xi_\mathrm{max} - 2)} = \varepsilon^2
\end{equation}
for $\xi_\mathrm{max}$, and then computing the value $\lambda^2$ from Eq.\ (\ref{ximax}). We denote the value of $\lambda$ obtained this way as $\lambda_c(\varepsilon)$. The physically relevant root of Eq.\ (\ref{forlambdac}) can be written as
\begin{widetext}
\[ \xi_\mathrm{max}(\varepsilon) = \begin{cases}
\frac{2 \varepsilon \sqrt{4 \varepsilon^2-3} \sin \left(\frac{1}{6} \left(2 \cos^{-1}\left(\frac{-16 \varepsilon^4 + 45 \varepsilon^2-27}{2 \varepsilon (4\varepsilon^2-3)^{3/2}}\right)+\pi \right)\right)+2 \varepsilon^2-3}{3 (\varepsilon^2-1)},
& 1 < \varepsilon^2 \le \frac{3}{32}(15 + \sqrt{33}), \\
\frac{2 \varepsilon \sqrt{4 \varepsilon^2-3} \cos \left(\frac{1}{3} \cos ^{-1}\left(\frac{16 \varepsilon^4-45 \varepsilon^2+27}{2 \varepsilon (4 \varepsilon^2-3)^{3/2}}\right)\right)+2 \varepsilon^2-3}{3 (\varepsilon^2-1)}, & \varepsilon^2 > \frac{3}{32}(15 + \sqrt{33}). \end{cases} \]
\end{widetext}
The above solution drops from $\xi_\mathrm{max} = \frac{3 + \sqrt{5}}{2}$ for $\varepsilon = 1$ to $\xi_\mathrm{max} = 2$ for $\varepsilon \to \infty$. While the above formula is written in explicitly real terms, in numerical applications it might be more convenient to use simpler complex-valued expressions. The function $\lambda_c(\varepsilon)$ is then computed as \[\lambda_c(\varepsilon) = \frac{\xi_\mathrm{max}(\varepsilon)}{\sqrt{\xi_\mathrm{max}(\varepsilon) - 2}}. \]

All particles that can reach infinity have $\varepsilon^2 \ge 1$. Particles with $\varepsilon^2 \ge 1$ and $\lambda < \lambda_c(\varepsilon)$ are absorbed by the black hole. Moreover, these are the only constraints on the family of particles which travel from infinity and fall into the black hole.

The description of the particles that travel from infinity with sufficiently high angular momentum and are scattered back to infinity is more complex. First, one observes that that the minimal energy $\varepsilon$ of a scattered particle at a given radius $\xi$ is given by
\begin{equation}
    \varepsilon_\mathrm{min} = \begin{cases} \infty, & \xi \leq 2 ,\\
    \sqrt{\frac{(\xi - 1)^3}{\xi^2 (\xi - 2)}}, & 2 < \xi \leq \frac{3 + \sqrt{5}}{2}, \\
    1, & \xi > \frac{3 + \sqrt{5}}{2}.  \end{cases}
\end{equation}
Note that $\xi = 2$ corresponds to the radius of the photon sphere. Consequently, no scattered particles can be found below the photon sphere. Next, since the motion of any particle is only allowed in a region where $\varepsilon^2 \ge U_\lambda(\xi)$, i.e,
\[ \varepsilon^2 \ge \frac{(\xi - 1)^2}{\xi^2} \left( 1 + \frac{\lambda^2}{\xi^2} \right), \]
the maximal allowed angular momentum $\lambda_\mathrm{max}(\varepsilon,\xi)$ is given by
\[ \lambda_\mathrm{max}(\varepsilon,\xi) = \xi \sqrt{\frac{\xi^2 \varepsilon^2}{(\xi - 1)^2} - 1}. \]
Scattered particles occupy the range in the phase space specified as $\varepsilon_\mathrm{min} \leq \varepsilon$, $\lambda_c(\varepsilon) < \lambda \leq \lambda_\mathrm{max}$.

\subsubsection{General Reissner-Nordstr\"{o}m metrics with $0 \le q \le 1$}

In the general case with $0 \le q \le 1$ we have
\[ N = \left( 1 - \frac{2}{\xi} + \frac{q^2}{\xi^2} \right) \]
and
\begin{equation}
\label{pot_gen_rn}
U_\lambda(\xi) = \left( 1 - \frac{2}{\xi} + \frac{q^2}{\xi^2} \right) \left( 1 + \frac{\lambda^2}{\xi^2} \right).
\end{equation}
The derivative $dU_\lambda/d\xi$ reads
\[ \frac{dU_\lambda}{d\xi} = \frac{2 \left[ \xi^3 - (q^2 + \lambda^2) \xi^2 + 3 \lambda^2 \xi - 2 q^2 \lambda^2 \right]}{\xi^5}. \]
Consequently, the locations of the extrema of $U_\lambda$ can be easily computed using Cardano's formulas. 

We proceed further using the same trick, as before. Note that at the extremum of the potential $U_\lambda$, we have
\begin{equation}
\label{lambda_gen_rn}
\lambda^2 = \frac{\xi^2(\xi - q^2)}{\xi^2 - 3 \xi + 2 q^2}.
\end{equation}
Inserting this expression in Eq.\ (\ref{pot_gen_rn}) we get
\begin{equation}
\label{ulambda_gen_rn}
U_\lambda = \frac{(\xi^2 - 2 \xi + q^2)^2}{\xi^2 (\xi^2 - 3 \xi + 2 q^2)} = \frac{(\xi-\xi_+)^2(\xi-\xi_-)^2}{\xi^2 (\xi-\xi_\mathrm{ph+})(\xi-\xi_\mathrm{ph-})},
\end{equation}
which is valid at the extrema of $U_\lambda$. In the last expression $\xi_\mathrm{ph\pm} = \frac{1}{2} \left(3 \pm \sqrt{9 - 8 q^2} \right)$ are the radii of the circular photon orbits. The radius of the outer photon sphere corresponds to $\xi = \xi_\mathrm{ph+}$. Note that $\xi^2 - 3\xi + 2 q^2 > 0$ for $\xi > \xi_\mathrm{ph+}$. Also $U_\lambda \to \infty$ for $\xi \to \xi_\mathrm{ph+}$ (the other root of $\xi^2 - 3\xi + 2 q^2$, i.e., $\xi = \xi_\mathrm{ph-}$ is always located below the black-hole horizon: we have $\xi_\mathrm{ph-} \le \xi_+$ for $0 \le q \le 1$). Consequently, the corresponding expression for $\varepsilon_\mathrm{min}$ reads
\begin{equation}
    \varepsilon_\mathrm{min} = \begin{cases} \infty, & \xi \leq \frac{1}{2} \left( 3 + \sqrt{9 - 8q^2} \right) ,\\
    \frac{\xi^2 - 2 \xi + q^2}{\xi \sqrt{\xi^2 - 3\xi + 2 q^2}}, & \frac{1}{2} \left( 3 + \sqrt{9 - 8q^2} \right) < \xi \le X(q), \\
    1, & \xi > X(q).  \end{cases}
\end{equation}
Here $X(q)$ is a unique root of the equation $\xi^3 - 4 \xi^2 + 4q^2 \xi - q^4 = 0$ satisfying $ (3 + \sqrt{5})/2 \le X(q) \le 4$. It can be written as
\begin{widetext}
\[ X(q) = \frac{4}{3} \left(\sqrt{4-3 q^2} \cos \left\{\frac{1}{3} \cos ^{-1}\left[\frac{\frac{27
   q^4}{16}-9 q^2+8}{\left(4-3 q^2\right)^{3/2}}\right]\right\}+1\right). \]
\end{widetext}

Note that, quite generally, the condition $\varepsilon^2 \ge U_\lambda(\xi)$ is equivalent to 
\[ \lambda < \xi \sqrt{\frac{\varepsilon^2}{N} - 1}. \]
In deriving the above inequality, one only assumes that $N > 0$. Consequently, for the general Reissner-Nordstr\"{o}m metric with $0 \le q \le 1$ we have
\[ \lambda_\mathrm{max}(\varepsilon,\xi) = \xi \sqrt{\frac{\xi^2 \varepsilon^2}{\xi^2 -2 \xi + q^2} - 1}. \]

The value $\lambda_c(\varepsilon)$ can be also computed analytically, but the corresponding formulas are lengthy, and of little practical importance (setting $U_\lambda = \varepsilon^2$ in Eq.\ (\ref{ulambda_gen_rn}) yields a quartic equation for $\xi$; the appropriate solution has to be inserted in Eq.\ (\ref{lambda_gen_rn})). In practice, one can always compute $\lambda_c(\varepsilon)$ numerically.

To recapitulate this section, let us note that the range of integration of the momentum-space integrals derived in Sec.\ \ref{momentumintegrals} is effectively limited by three functions: $\lambda_c(\varepsilon)$, $\varepsilon_\mathrm{min}(\xi)$, and $\lambda_\mathrm{max}(\varepsilon,\xi)$. The first one, $\lambda_c(\varepsilon)$, is defined as the angular momentum parameter $\lambda$ for which the value of the effective potential $U_\lambda$ at its local maximum equals precisely $\varepsilon^2$. The second, $\varepsilon_\mathrm{min}(\xi)$, gives the minimal energy $\varepsilon$ of a scattered particle at a given radius $\xi$. The function $\lambda_\mathrm{max}(\varepsilon,\xi)$ yields an upper bound on the angular momentum $\lambda$ of a scattered particle with the energy $\varepsilon$ at the radius $\xi$. The range of the phase-space occupied by absorbed particles is limited by $\varepsilon \ge 1$ and $0 < \lambda < \lambda_c(\varepsilon)$. The range corresponding to scattered particles is given by $\varepsilon_\mathrm{min} \le \varepsilon$, $\lambda_c(\varepsilon) < \lambda \le \lambda_\mathrm{max}$. A detailed proof of this characterization is given in \cite{Olivier} for the Schwarzschild metric. 

\section{Accretion of the gas in thermal equilibrium at infinity}
\label{sec:gasinequilibrium}

We will now compute the momenum integrals of Sec.\ \ref{momentumintegrals}, assuming the Maxwell-J\"{u}ttner distribution function (\ref{juttnerPQ}). This yields
\begin{equation}
\label{juttnercapitalf}
\mathcal F_n(\varepsilon, \lambda) = m^n \alpha e^{-z \varepsilon}
\end{equation}
for the distribution function (\ref{massmoment}). Integral quantities $J_\mu$ and $T_{\mu \nu}$ can be now divided into two parts, corresponding to absorbed and scattered particles. For the particle current density $J_\mu = J_\mu^{\mathrm{(abs)}} + J_\mu^{\mathrm{(scat)}}$ we have
\begin{widetext}
\begin{eqnarray*}
J_t^{\mathrm{(abs)}}  & = & - \frac{2 \pi \alpha m^4}{\xi^2} \int_1^\infty d\varepsilon e^{-z \varepsilon} \varepsilon \int_0^{\lambda_c} d \lambda \frac{\lambda}{\sqrt{\varepsilon^2 - U_\lambda(\xi)}} = - \frac{2 \pi \alpha m^4}{\xi^2} \int_1^\infty d\varepsilon e^{-z \varepsilon} \varepsilon \frac{\lambda_c^2}{s_{\lambda_c} + s_0}, \\
J_r^{\mathrm{(abs)}} & = & \frac{2 \pi \alpha m^4}{\xi^2} \int_1^\infty d \varepsilon e^{-z \varepsilon} \int_0^{\lambda_c} d \lambda \pi_{\xi-}  \frac{\lambda}{\sqrt{\varepsilon^2 - U_\lambda (\xi) }} = \frac{2 \pi \alpha m^4}{\xi^2 N} \int_1^\infty d \varepsilon e^{-z \varepsilon} \lambda_c^2 \left[ \frac{(1 - N \eta) \varepsilon}{s_{\lambda_c} + s_0} - \frac{1}{2} \right].
\end{eqnarray*}
Here we choose $\pi_\xi = \pi_{\xi-}$, which corresponds to ingoing particles. The formulas for $J_\mu^\mathrm{(scat)}$ are slightly more complex:
\begin{eqnarray*}
J_t^{\mathrm{(scat)}} & = & - \frac{4 \pi \alpha m^4}{\xi^2} \int_{\varepsilon_\mathrm{min}}^\infty d\varepsilon e^{-z \varepsilon} \varepsilon \int_{\lambda_c}^{\lambda_\mathrm{max}} d \lambda \frac{\lambda}{\sqrt{\varepsilon^2 - U_\lambda (\xi) }} = - \frac{4 \pi \alpha m^4}{\xi^2} \int_{\varepsilon_\mathrm{min}}^\infty d\varepsilon e^{-z \varepsilon} \varepsilon \frac{\lambda_\mathrm{max}^2 - \lambda_c^2}{s_{\lambda_c} + s_{\lambda_\mathrm{max}}}, \\
J_r^{\mathrm{(scat)}} & = & \frac{2 \pi \alpha m^4}{\xi^2} \sum_\pm \int_{\varepsilon_\mathrm{min}}^\infty d \varepsilon e^{-z \varepsilon} \int_{\lambda_c}^{\lambda_\mathrm{max}} d \lambda   \frac{\lambda \pi_{\xi \pm}}{\sqrt{\varepsilon^2 - U_\lambda (\xi) }}  = \frac{4 \pi \alpha m^4}{\xi^2} \frac{(1 - N \eta)}{N} \int_{\varepsilon_\mathrm{min}}^\infty d \varepsilon e^{-z \varepsilon} \varepsilon \frac{\lambda_\mathrm{max}^2 - \lambda_c^2}{s_{\lambda_c} + s_{\lambda_\mathrm{max}}}.
\end{eqnarray*}
\end{widetext}
Here the additional overall factor 2 stems from the fact that both ingoing and outgoing particles are taken into account.

In the above formulas and in the following text we omit the arguments of the functions $\lambda_c$, $\lambda_\mathrm{max}$, $s_0$, $s_{\lambda_c}$, $s_{\lambda_\mathrm{max}}$, and $\varepsilon_\mathrm{min}$. We recall that $\lambda_\mathrm{max} = \lambda_\mathrm{max}(\varepsilon,\xi)$, $s_0 = s_0(\varepsilon,\xi)$, $s_{\lambda_c} = s_{\lambda_c}(\varepsilon,\xi)$, $s_{\lambda_\mathrm{max}} = s_{\lambda_\mathrm{max}}(\varepsilon,\xi)$ depend on both $\varepsilon$ and $\xi$. The energy $\varepsilon_\mathrm{min} = \varepsilon_\mathrm{min}(\xi)$ is a function of $\xi$, while $\lambda_c = \lambda_c(\varepsilon)$ is a function of $\varepsilon$ only.

Note that $J_r^\mathrm{(scat)} = - \frac{g^{tr}}{g^{rr}}J_t^\mathrm{(scat)}$. Accordingly, $J^{r \mathrm{(scat)}} = 0$. It follows that only $J_\mu^\mathrm{(abs)}$ contributes to the mass accretion rate $\dot M$. A simple calculation yields
\begin{eqnarray}
\dot M & = & - 4 \pi m r^2 J^r = 8 \pi^2 m M^2 \int_1^\infty d \varepsilon \int_0^{\lambda_c} d \lambda \mathcal F_4 (\varepsilon, \lambda) \lambda \nonumber \\
& = & 4 \pi^2 M^2 m^5 \alpha \int_1^\infty d \varepsilon e^{-z \varepsilon} \lambda_c^2.
\label{mdotcomputed}
\end{eqnarray}

The analysis of the energy-momentum tensor is especially interesting in terms of its spectral properties. For perfect fluids, which can serve as a reference for the more complex case of the Vlasov gas, the energy momentum tensor reads
\begin{equation}
\label{tmunu_perf_fluid}
T\indices{^\mu_\nu} = (\rho + p)u^\mu u_\nu + p \delta^\mu_\nu,
\end{equation}
where $u^\mu$ denotes the four velocity of the fluid ($u^\mu u_\mu = -1$), $\rho$ is the energy density, and $p$ denotes the pressure. It is easy to see that $u^\mu$ is an eigenvector of $T\indices{^\mu_\nu}$ corresponding to the eigenvalue $- \rho$. On the other hand any non-zero vector $k^\mu$ orthogonal to the four-velocity ($k^\mu u_\mu = 0$) is also an eigenvector corresponding to the eigenvalue $p$. Consequently, the pressure $p$ is a three-fold degenerate eigenvalue of $T\indices{^\mu_\nu}$.

Solving the eigenvalue problem for $T\indices{^\mu_\nu}$ gives a possibility to compare the properties of the Vlasov gas with those of the perfect fluid. In particular, we will see that the eigenvalue corresponding to the pressure is no longer three-fold degenerate. Instead, we obtain two-fold degeneracy due to the assumed spherical symmetry. For perfect fluids, the (conserved) particle current density can be expressed as $J^\mu = n u^\mu$, where $n$ denotes the particle density. In other words, $J^\mu$ is proportional to the four-velocity $u^\mu$, i.e., to the timelike eigenvector of $T\indices{^\mu_\nu}$. For the Vlasov gas this does not have to be the case.

Raising the first index in Eqs.\ (\ref{Tthetatheta}--\ref{Trr}) we get
\begin{widetext}
\begin{eqnarray*}
T\indices{^t_t} & = & - \frac{2 \pi}{\xi^2 N} \int \mathcal F_5(\varepsilon, \lambda) \varepsilon \lambda \left[ \frac{\varepsilon}{\sqrt{\varepsilon^2 - U_\lambda(\xi)}} + \epsilon(\pi_\xi) (1 - N \eta) \right] d\varepsilon d\lambda, \\
T\indices{^t_r} & = & \frac{2 \pi}{\xi^2 N} \int \mathcal F_5(\varepsilon, \lambda) \lambda \pi_\xi \left[ \frac{\varepsilon}{\sqrt{\varepsilon^2 - U_\lambda(\xi)}} + \epsilon(\pi_\xi) (1 - N \eta) \right] d\varepsilon d\lambda, \\
T\indices{^r_t} & = & - \frac{2 \pi}{\xi^2} \int \epsilon(\pi_\xi) \mathcal F_5(\varepsilon, \lambda) \varepsilon \lambda d\varepsilon d\lambda, \\
T\indices{^r_r} & = &  \frac{2\pi}{\xi^2} \int \epsilon(\pi_\xi) \mathcal F_5(\varepsilon, \lambda) \lambda \pi_\xi d \varepsilon d\lambda\\
T\indices{^\theta_\theta} & = & T\indices{^\varphi_\varphi} = \frac{\pi}{\xi^4} \int  \frac{\mathcal F_5(\varepsilon, \lambda) \lambda^3}{\sqrt{\varepsilon^2 - U_\lambda(\xi)}} d \varepsilon d\lambda.
\end{eqnarray*}

It turns out that the spectral properties of $T\indices{^\mu_\nu}$ differ significantly between the contributions related with absorbed trajectories ${T_\mathrm{(abs)}}\indices{^\mu_\nu}$ and scattered trajectories ${T_\mathrm{(scat)}}\indices{^\mu_\nu}$. For scattered trajectories we have
\begin{eqnarray*}
{T_\mathrm{(scat)}}\indices{^t_t} & = & - \frac{4 \pi}{\xi^2 N} \int  \frac{ \mathcal F_5(\varepsilon, \lambda) \lambda \varepsilon^2 }{\sqrt{\varepsilon^2 - U_\lambda(\xi)}} d\varepsilon d\lambda, \\
{T_\mathrm{(scat)}}\indices{^t_r} & = & \frac{4 \pi (1 - N \eta)}{\xi^2 N^2} \int \mathcal F_5(\varepsilon, \lambda) \lambda \left[ \frac{\varepsilon^2}{\sqrt{\varepsilon^2 - U_\lambda(\xi)}} + \sqrt{\varepsilon^2 - U_\lambda(\xi)} \right] d\varepsilon d\lambda, \\
{T_\mathrm{(scat)}}\indices{^r_t} & = & 0, \\
{T_\mathrm{(scat)}}\indices{^r_r} & = & \frac{4\pi}{\xi^2 N} \int \mathcal F_5(\varepsilon, \lambda) \lambda \sqrt{\varepsilon^2 - U_\lambda(\xi)} d \varepsilon d\lambda, \\
{T_\mathrm{(scat)}}\indices{^\theta_\theta} & = & {T_\mathrm{(scat)}}\indices{^\varphi_\varphi} = \frac{2\pi}{\xi^4} \int \frac{\mathcal F_5(\varepsilon, \lambda) \lambda^3}{\sqrt{\varepsilon^2 - U_\lambda(\xi)}} d \varepsilon d\lambda.
\end{eqnarray*}
\end{widetext}
It can be checked that the orthogonal frame $(\partial_t, - (g_{tr}/g_{tt}) \partial_t + \partial_r,  \partial_\theta, \partial_\varphi)$ consists of eigenvectors of ${T_\mathrm{(scat)}}\indices{^\mu_\nu}$. The corresponding eigenvalues can be obtained as follows: The eigenvalue corresponding to the timelike eigenvector $\partial_t$ reads
\[ \varrho^\mathrm{(scat)} = \frac{4 \pi}{\xi^2 N} \int  \frac{\mathcal F_5(\varepsilon, \lambda) \lambda \varepsilon^2}{\sqrt{\varepsilon^2 - U_\lambda(\xi)}} d \varepsilon d \lambda, \]
and it can be identified with the energy-density. Note that this also means that, similarly to perfect fluids, for scattered particles the eigenvector corresponding to the energy density is proportional to $J^\mu$. The eigenvalue corresponding to the radial eigenvector $- (g_{tr}/g_{tt}) \partial_t + \partial_r$ is
\[ p_\mathrm{rad}^\mathrm{(scat)} = \frac{4 \pi}{\xi^2 N} \int \mathcal F_5(\varepsilon, \lambda) \lambda \sqrt{\varepsilon^2 - U_\lambda(\xi)} d \varepsilon d \lambda. \]
It can be understood as the radial pressure. The eigenvalue corresponding to eigenvectors $\partial_\theta$ and $\partial_\varphi$ is two-fold degenerate. It reads
\begin{equation}
\label{ptangeneral}
p_\mathrm{tan}^\mathrm{(scat)} = \frac{2 \pi}{\xi^4} \int  \frac{\mathcal F_5(\varepsilon, \lambda) \lambda^3}{\sqrt{\varepsilon^2 - U_\lambda(\xi)}} d \varepsilon d \lambda
 \end{equation}
and will be referred to as the tangential pressure. For $\mathcal F_5(\varepsilon, \lambda)$ given by Eq. (\ref{juttnercapitalf}) we get
\begin{widetext}
\begin{eqnarray*}
\rho^\mathrm{(scat)} & = & \frac{4 \pi \alpha m^5}{\xi^2 N} \int_{\varepsilon_\mathrm{min}}^\infty d \varepsilon e^{-z \varepsilon} \varepsilon^2 \frac{\lambda_\mathrm{max}^2 - \lambda_c^2}{s_{\lambda_c} + s_{\lambda_\mathrm{max}}}, \\
p_\mathrm{rad}^\mathrm{(scat)} & = & \frac{4 \pi \alpha m^5}{3 N^2} \int_{\varepsilon_\mathrm{min}}^\infty d \varepsilon e^{-z \varepsilon} \left( s_{\lambda_c}^3 - s_{\lambda_\mathrm{max}}^3 \right), \\
p_\mathrm{tan}^\mathrm{(scat)} & = & \frac{2 \pi \alpha m^5}{3 \xi^4} \int_{\varepsilon_\mathrm{min}}^\infty d \varepsilon e^{-z \varepsilon} \frac{\left( \lambda_\mathrm{max}^2 - \lambda_c^2 \right) \left[ \lambda^2_\mathrm{max} \left( 2 s_{\lambda_c} + s_{\lambda_\mathrm{max}} \right) + \lambda_c^2 \left( s_{\lambda_c} + 2 s_{\lambda_\mathrm{max}} \right) \right]}{\left( s_{\lambda_c} + s_{\lambda_\mathrm{max}} \right)^2} .
\end{eqnarray*}
In the following, we will also need explicit expressions for ${T_\mathrm{(scat)}}\indices{^\mu_\nu}$ with $\mathcal F_5(\varepsilon, \lambda)$ given by Eq.\ (\ref{juttnercapitalf}). These are
\begin{eqnarray*}
{T_\mathrm{(scat)}}\indices{^t_t} & = & - \frac{4 \pi \alpha m^5}{\xi^2 N} \int_{\varepsilon_\mathrm{min}}^\infty d \varepsilon e^{-z \varepsilon}  \varepsilon^2 \frac{\lambda_\mathrm{max}^2 - \lambda_c^2}{s_{\lambda_c} + s_{\lambda_\mathrm{max}}}, \\
{T_\mathrm{(scat)}}\indices{^t_r} & = & \frac{4 \pi \alpha m^5 (1 - N \eta)}{\xi^2 N^2} \int_{\varepsilon_\mathrm{min}}^\infty d \varepsilon e^{-z \varepsilon} \left[ \varepsilon^2 \frac{\lambda_\mathrm{max}^2 - \lambda_c^2}{s_{\lambda_c} + s_{\lambda_\mathrm{max}}} + \frac{\xi^2}{3 N} \left(  s_{\lambda_c}^3 - s_{\lambda_\mathrm{max}}^3 \right) \right], \\
{T_\mathrm{(scat)}}\indices{^r_t} & = & 0, \\
{T_\mathrm{(scat)}}\indices{^r_r} & = & \frac{4\pi \alpha m^5}{3 N^2} \int_{\varepsilon_\mathrm{min}}^\infty  d\varepsilon e^{-z \varepsilon} \left( s_{\lambda_c}^3 - s_{\lambda_\mathrm{max}}^3 \right), \\
{T_\mathrm{(scat)}}\indices{^\theta_\theta} & = & {T_\mathrm{(scat)}}\indices{^\varphi_\varphi} = \frac{2 \pi \alpha m^5}{3 \xi^4} \int_{\varepsilon_\mathrm{min}}^\infty d \varepsilon e^{-z \varepsilon} \frac{\lambda_\mathrm{max}^2 - \lambda_c^2}{\left( s_{\lambda_c} + s_{\lambda_\mathrm{max}} \right)^2} \left[ \lambda^2_\mathrm{max} \left( 2 s_{\lambda_c} + s_{\lambda_\mathrm{max}} \right) + \lambda_c^2 \left( s_{\lambda_c} + 2 s_{\lambda_\mathrm{max}} \right) \right].
\end{eqnarray*}
\end{widetext}

For absorbed trajectories the situation is different. The tangential pressure, i.e., the eigenvalue corresponding to the eigenvectors $\partial_\theta$ and $\partial_\varphi$, can be written as
\[ p_\mathrm{tan}^\mathrm{(abs)} = \frac{2 \pi}{\xi^4} \int \frac{\mathcal F_5(\varepsilon, \lambda) \lambda^3}{\sqrt{\varepsilon^2 - U_\lambda(\xi)}} d \varepsilon d \lambda, \]
that is in the same form as Eq.\ (\ref{ptangeneral}). On the other hand the  the formulas for $\rho^\mathrm{(abs)}$ and $p_\mathrm{rad}^\mathrm{(abs)}$ are different. Moreover, $\partial_t$ and $- (g_{tr}/g_{tt}) \partial_t + \partial_r$ are no longer the appropriate eigenvectors. Instead, the timelike eigenvector corresponding to the energy density $\rho^\textrm{(abs)}$ is inclined at some angle with respect to $\partial_t$.

For absorbed trajectories, the components of the energy-momentum tensor can be written as
\begin{widetext}
\begin{eqnarray*}
{T_\mathrm{(abs)}}\indices{^t_t} & = & - \frac{2 \pi}{\xi^2 N} \int \mathcal F_5(\varepsilon, \lambda) \varepsilon \lambda \left[ \frac{\varepsilon}{\sqrt{\varepsilon^2 - U_\lambda(\xi)}} - (1 - N \eta) \right] d\varepsilon d\lambda, \\
{T_\mathrm{(abs)}}\indices{^t_r} & = & \frac{2 \pi}{\xi^2 N} \int \mathcal F_5(\varepsilon, \lambda) \lambda \pi_{\xi -} \left[ \frac{\varepsilon}{\sqrt{\varepsilon^2 - U_\lambda(\xi)}} - (1 - N \eta) \right] d\varepsilon d\lambda, \\
{T_\mathrm{(abs)}}\indices{^r_t} & = & \frac{2 \pi}{\xi^2} \int \mathcal F_5(\varepsilon, \lambda) \varepsilon \lambda d\varepsilon d\lambda, \\
{T_\mathrm{(abs)}}\indices{^r_r} & = & - \frac{2\pi}{\xi^2} \int \mathcal F_5(\varepsilon, \lambda) \lambda \pi_{\xi-} d \varepsilon d\lambda\\
{T_\mathrm{(abs)}}\indices{^\theta_\theta} & = & {T_\mathrm{(abs)}}\indices{^\varphi_\varphi} = \frac{\pi}{\xi^4} \int  \frac{\mathcal F_5(\varepsilon, \lambda) \lambda^3}{\sqrt{\varepsilon^2 - U_\lambda(\xi)}} d \varepsilon d\lambda.
\end{eqnarray*}
For $\mathcal F_5(\varepsilon, \lambda)$ given by Eq.\ (\ref{juttnercapitalf}) the above integrals can be evaluated as
\begin{eqnarray*}
{T_\mathrm{(abs)}}\indices{^t_t} & = & - \frac{2 \pi \alpha m^5}{\xi^2 N} \int_1^\infty d \varepsilon e^{-z \varepsilon} \varepsilon \lambda_c^2 \left[ \frac{\varepsilon}{s_{\lambda_c} + s_0} - \frac{(1 - N \eta)}{2} \right],\\
{T_\mathrm{(abs)}}\indices{^t_r} & = & \frac{2 \pi \alpha m^5}{\xi^2 N^2} \int_1^\infty d \varepsilon e^{-z \varepsilon} \left\{ \frac{(1 - N \eta) \varepsilon^2 \lambda_c^2}{s_{\lambda_c} + s_0} - \frac{1}{2} \left[ 1 + (1 - N \eta)^2 \right] \varepsilon \lambda_c^2 + \frac{(1 - N \eta) \xi^2}{3 N} \left( s_0^3 - s_{\lambda_c}^3 \right) \right\}, \\
{T_\mathrm{(abs)}}\indices{^r_t} & = & \frac{\pi \alpha m^5}{\xi^2} \int_1^\infty d \varepsilon e^{-z \varepsilon} \varepsilon \lambda_c^2, \\
{T_\mathrm{(abs)}}\indices{^r_r} & = & - \frac{2 \pi \alpha m^5}{\xi^2 N} \left[ \frac{(1 - N \eta)}{2} \int_1^\infty d \varepsilon e^{-z \varepsilon} \varepsilon \lambda_c^2 + \frac{\xi^2}{3 N} \int_1^\infty d \varepsilon e^{- z \varepsilon} \left( s_{\lambda_c}^3 - s_0^3 \right)  \right], \\
{T_\mathrm{(abs)}}\indices{^\theta_\theta} & = & {T_\mathrm{(abs)}}\indices{^\varphi_\varphi} = \frac{\pi \alpha m^5}{3 \xi^4} \int_1^\infty d \varepsilon e^{-z \varepsilon} \lambda_c^4 \frac{2 s_0 + s_{\lambda_c}}{(s_0 + s_{\lambda_c})^2}.
\end{eqnarray*}

While the above expressions are relatively compact, they suffer from the occurrence of expressions that would become indeterminate (of the form $0/0$) at the horizon. This drawback can be removed by rewriting them in the form
\begin{subequations}
\label{tmunuabsregular}
\begin{eqnarray}
{T_\mathrm{(abs)}}\indices{^t_t} & = & - \frac{2 \pi \alpha m^5}{\xi^2} \int_1^\infty d \varepsilon e^{-z \varepsilon} \frac{\varepsilon \lambda_c^2}{s_0 + s_{\lambda_c}} \left[ \eta \varepsilon + \frac{(1 - N \eta)}{2} X \right] ,\\
{T_\mathrm{(abs)}}\indices{^t_r} & = &  \frac{2 \pi \alpha m^5}{\xi^2} \int_1^\infty d \varepsilon e^{-z \varepsilon} \left\{ \frac{\lambda_c^2}{s_0 + s_{\lambda_c}} \left[ \eta \varepsilon + \frac{(1 - N \eta)}{2} X \right] \left( - \eta \varepsilon + \frac{1}{2} X \right) - \frac{(1 - N \eta) \xi^2}{12} Y^3 \right\}, \\
  {T_\mathrm{(abs)}}\indices{^r_t} & = & \frac{\pi \alpha m^5}{\xi^2} \int_1^\infty d \varepsilon e^{-z \varepsilon} \varepsilon \lambda_c^2, \\
{T_\mathrm{(abs)}}\indices{^r_r} & = &  -\frac{2 \pi \alpha m^5}{\xi^2} \int_1^\infty d \varepsilon e^{-z \varepsilon} \lambda_c^2 \left( - \frac{\eta \varepsilon}{2} + \frac{1}{4}X + \frac{\xi^2 N}{12 \lambda_c^2} Y^3 \right), \\
{T_\mathrm{(abs)}}\indices{^\theta_\theta} & = & {T_\mathrm{(abs)}}\indices{^\varphi_\varphi} = \frac{\pi \alpha m^5}{3 \xi^4} \int_1^\infty d \varepsilon e^{-z \varepsilon} \lambda_c^4 \frac{2 s_0 + s_{\lambda_c}}{(s_0 + s_{\lambda_c})^2},
\end{eqnarray}
\end{subequations}
\end{widetext}
where
\[X = \frac{1}{\varepsilon + s_0} + \frac{1 + \frac{\lambda_c^2}{\xi^2}}{\varepsilon + s_{\lambda_c}}, \quad Y = \frac{1}{\varepsilon + s_0} - \frac{1 + \frac{\lambda_c^2}{\xi^2}}{\varepsilon + s_{\lambda_c}}. \]
Using expressions that are manifestly regular at the horizon is important both from a purely theoretical perspective and also numerically, as it helps to avoid numerical errors in evaluating the above expressions in the vicinity of the horizon. (Note that this is irrelevant with respect to ${T_\mathrm{(scat)}}\indices{^\mu_\nu}$, since these components vanish identically in the vicinity of the horizon.) Taking the limit of expressions (\ref{tmunuabsregular}) as $\xi \to \xi_+ = 1 + \sqrt{1 - q^2}$ (the horizon) we get
\begin{widetext}
\begin{eqnarray*}
{T_\mathrm{(abs)}}\indices{^t_t}\Bigr|_{\substack{\xi=\xi_+}} & = & - \frac{2 \pi \alpha m^5}{\xi_+^2} \int_1^\infty d \varepsilon ^{-z \varepsilon} \left( \frac{\lambda_c^2\eta\varepsilon}{2} + \frac{\lambda_c^2}{4\varepsilon} + \frac{\lambda_c^4}{8\varepsilon\xi_+^2} \right) ,\\
{T_\mathrm{(abs)}}\indices{^t_r}\Bigr|_{\substack{\xi=\xi_+}}  & = &  \frac{2 \pi \alpha m^5}{\xi_+^2} \int_1^\infty d \varepsilon e^{-z \varepsilon} \left[ -\frac{\lambda_c^2\eta^2\varepsilon}{2} + \frac{\lambda_c^2}{8\varepsilon^3} + \frac{\lambda_c^4}{8\varepsilon^3\xi_+^2}+ \frac{\lambda_c^6}{24\varepsilon^3\xi_+^4}\right], \\
{T_\mathrm{(abs)}}\indices{^r_t}\Bigr|_{\substack{\xi=\xi_+}} & = & \frac{\pi \alpha m^5}{\xi_+^2} \int_1^\infty d \varepsilon e^{-z \varepsilon} \varepsilon \lambda_c^2, \\
{T_\mathrm{(abs)}}\indices{^r_r}\Bigr|_{\substack{\xi=\xi_+}} & = &  \frac{2 \pi \alpha m^5}{\xi_+^2} \int_1^\infty d \varepsilon e^{-z \varepsilon}  \left(\frac{\lambda_c^2\eta\varepsilon}{2} - \frac{\lambda_c^2}{4\varepsilon} - \frac{\lambda_c^4}{8\varepsilon\xi_+^2}\right), \\
{T_\mathrm{(abs)}}\indices{^\theta_\theta}\Bigr|_{\substack{\xi=\xi_+}} & = & {T_\mathrm{(abs)}}\indices{^\varphi_\varphi}\Bigr|_{\substack{\xi=\xi_+}} = \frac{\pi \alpha m^5}{4 \xi_+^4} \int_1^\infty d \varepsilon e^{-z \varepsilon}  \frac{\lambda_c^4}{\varepsilon}.
\end{eqnarray*}
\end{widetext}
In evaluating the above limits one makes use of the fact that $N \to 0$, $N \eta \to 0$ (this is actually an assumption on $\eta$),
\[ X \to \frac{1}{2 \varepsilon} \left( 2 + \frac{\lambda_c^2}{\xi^2} \right), \quad Y \to - \frac{\lambda_c^2}{2 \varepsilon \xi^2}, \]
as $\xi \to \xi_+$.

The eigenvalues $\rho^\mathrm{(abs)}$, $p_\mathrm{rad}^\mathrm{(abs)}$, and $p_\mathrm{tan}^\mathrm{(abs)}$ of ${T_\mathrm{(abs)}}\indices{^\mu_\nu}$ can be computed by ``brute force''. A rather lengthy calculation allows one to express $\rho^\mathrm{(abs)}$, $p_\mathrm{rad}^\mathrm{(abs)}$ as
\begin{subequations}
\begin{eqnarray}
\rho^\mathrm{(abs)} & = &  -\frac{\pi \alpha m^5}{\xi^2} (A - \sqrt{N^2 B^2 + 2 B C}), \\
p_\mathrm{rad}^\mathrm{(abs)} & = & \frac{\pi \alpha m^5}{\xi^2} (A + \sqrt{N^2 B^2 + 2 B C}),
\end{eqnarray}
\end{subequations}
where
\begin{subequations}
\begin{eqnarray}
A & = & - \frac{1}{4} \int_1^\infty d \varepsilon e^{-z \varepsilon} \left[ \left( 1 + \frac{2 \varepsilon}{s_0 + s_{\lambda_c}} \right) \lambda_c^2 X + \frac{N \xi^2 Y^3}{3} \right], \\
B & = & \frac{1}{4} \int_1^\infty d \varepsilon e^{-z \varepsilon} \left( \frac{\lambda_c^2 X^2}{s_0 + s_{\lambda_c}} - \frac{\xi^2 Y^3}{3} \right), \\
C & = & \int_1^\infty d \varepsilon e^{-z \varepsilon} \varepsilon \lambda_c^2,
\label{integralC}
\end{eqnarray}
\end{subequations}
Although the components ${T_\mathrm{(abs)}}\indices{^\mu_\nu}$ depend on the choice of the time foliation (i.e., on $\eta$), the eigenvalues $\rho^\mathrm{(abs)}$, $p_\mathrm{rad}^\mathrm{(abs)}$, $p_\mathrm{tan}^\mathrm{(abs)}$ are independent of $\eta$. One can also check that asymptotically, i.e., for $\xi \to \infty$, both $p_\mathrm{rad}^\mathrm{(abs)}$ and $p_\mathrm{tan}^\mathrm{(abs)}$ converge to the same limit.

At the horizon, $\rho^\mathrm{(abs)}$, $p_\mathrm{rad}^\mathrm{(abs)}$, and $p_\mathrm{tan}^\mathrm{(abs)}$ are given by the following much simpler expressions:
\begin{subequations}
\begin{eqnarray}
\rho^\mathrm{(abs)}\Bigr|_{\xi=\xi_+} & = & \frac{\pi \alpha m^5}{\xi_+^2} \left( A^\prime +
\sqrt{ B^\prime C} \right), \\
p_\mathrm{rad}^\mathrm{(abs)}\Bigr|_{\xi=\xi_+} & = & -\frac{\pi \alpha m^5}{\xi_+^2} \left( A^\prime -
\sqrt{B^\prime C} \right), \\
p_\mathrm{tan}^\mathrm{(abs)}\Bigr|_{\xi=\xi_+} & = & \frac{\pi \alpha m^5}{4 \xi_+^4} \int_1^\infty  d \varepsilon e^{-z \varepsilon}  \frac{\lambda_c^4}{\varepsilon},
\end{eqnarray}
\end{subequations}
where
\begin{eqnarray*}
A^\prime & = & \int_1^\infty d \varepsilon e^{-z \varepsilon} \frac{\lambda_c^2}{2\varepsilon}\left( 1+\frac{\lambda_c^2}{2\xi_+^2} \right), \\
B^\prime & = & \int_1^\infty d \varepsilon e^{-z \varepsilon} \frac{\lambda_c^2}{4 \varepsilon^3} \left( 1 + \frac{\lambda_c^2}{\xi_+^2} + \frac{\lambda_c^4}{3 \xi_+^4} \right),
\end{eqnarray*}
and $C$ is given by Eq.\ (\ref{integralC}).

We should emphasise that both ${T_\mathrm{(abs)}}\indices{^\mu_\nu}$ and ${T_\mathrm{(scat)}}\indices{^\mu_\nu}$ are of limited physical interest separately. Instead, we are rather interested in the total energy-momentum tensor $T\indices{^\mu_\nu} = {T_\mathrm{(abs)}}\indices{^\mu_\nu} + {T_\mathrm{(scat)}}\indices{^\mu_\nu}$ and its eigenvalues. Although $p_\mathrm{tan} = p^\mathrm{(abs)}_\mathrm{tan} + p^\mathrm{(scat)}_\mathrm{tan}$, we have $p_\mathrm{rad} \neq p^\mathrm{(abs)}_\mathrm{rad} + p^\mathrm{(scat)}_\mathrm{rad}$ and $\rho \neq \rho^\mathrm{(abs)} + \rho^\mathrm{(scat)}$. This is because the eigenvectors corresponding to $p^\mathrm{(abs)}_\mathrm{rad}$ and $p^\mathrm{(scat)}_\mathrm{rad}$ (as well as to $\rho^\mathrm{(abs)}$ and $\rho^\mathrm{(scat)}$) are different, and the corresponding eigenvalues do not add. In practice, it is convenient to compute the eigenvalues of $T\indices{^\mu_\nu}$ numerically, and we do it in the next section. On the other hand, since below the photon sphere ${T_\mathrm{(scat)}}\indices{^\mu_\nu} = 0$, the above expressions for $\rho^\mathrm{(abs)}$, $p^\mathrm{(abs)}_\mathrm{rad}$, $p^\mathrm{(abs)}_\mathrm{tan}$ also give the total physical values of $\rho$, $p_\mathrm{rad}$, and $p_\mathrm{tan}$ in the vicinity of the horizon. In contrast to that, analytic expressions for $\rho^\mathrm{(scat)}$, $p^\mathrm{(scat)}_\mathrm{rad}$, $p^\mathrm{(scat)}_\mathrm{tan}$ are of theoretical interest only.
 
\section{Numerical results}
\label{sec:numerical}

\subsection{Particle current density, particle density, mass accretion rate}

\begin{figure}
    \centering
    \includegraphics[width=0.45\textwidth]{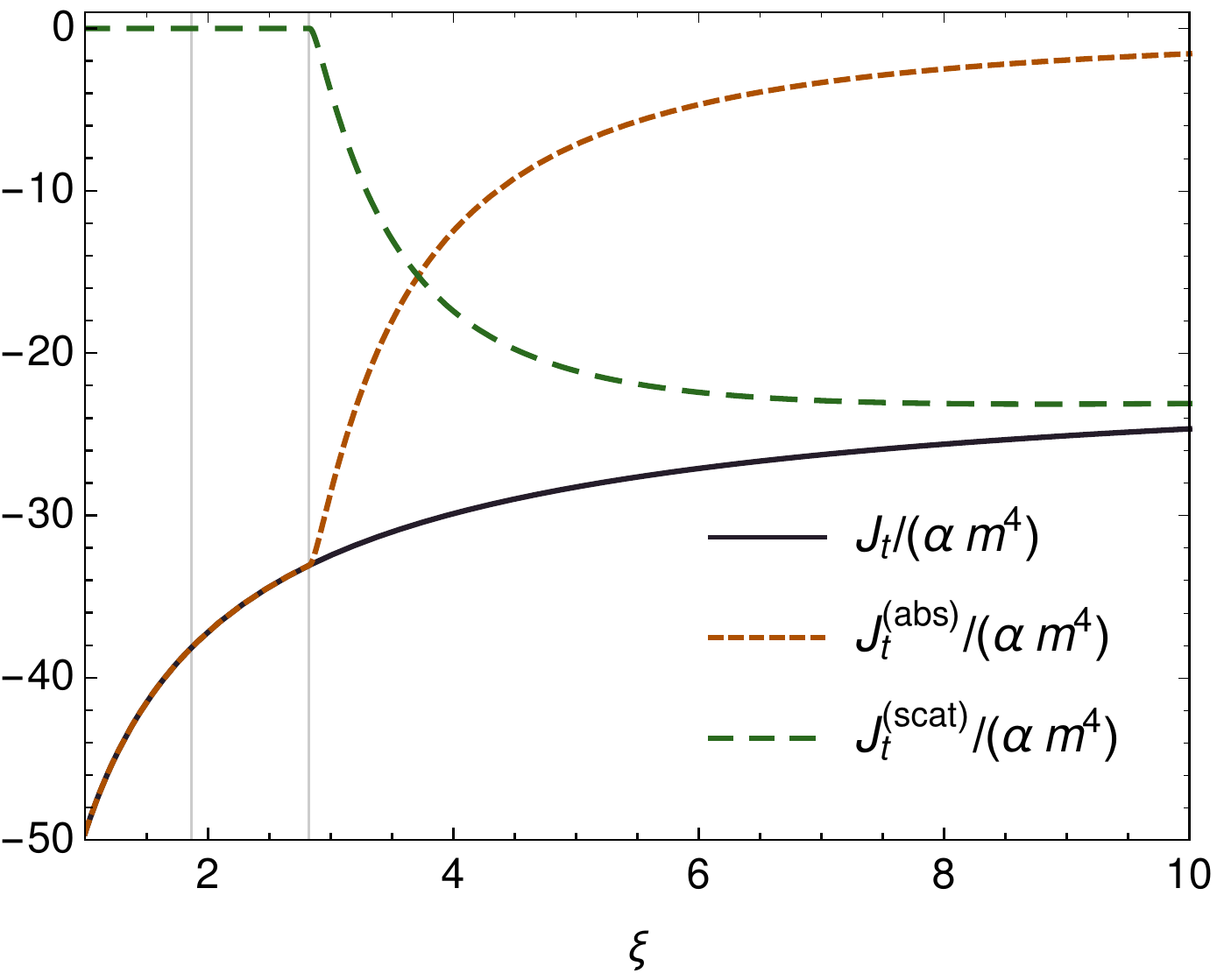}
    \caption{Sample graphs of $J_t/(\alpha m^4)$, $J^\mathrm{(abs)}_t/(\alpha m^4)$, and $J^\mathrm{(scat)}_t/(\alpha m^4)$. The charge parameter $q = 1/2$; $z = 1$. Vertical lines mark the locations of the black-hole horizon and the photon sphere.}
    \label{fig:jt}
\end{figure}

\begin{figure}
    \centering
    \includegraphics[width=0.45\textwidth]{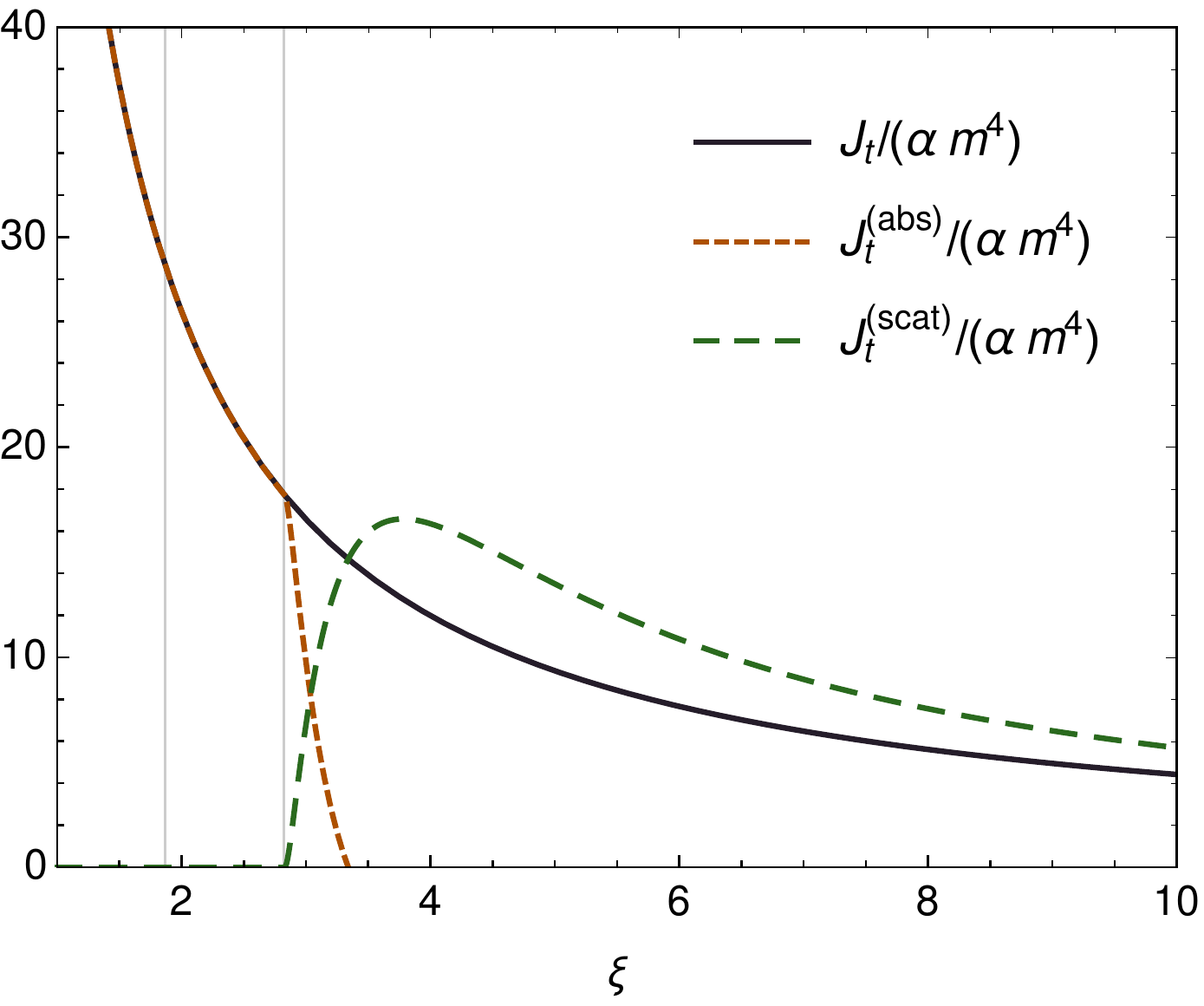}
    \caption{Sample graphs of $J_r/(\alpha m^4)$, $J^\mathrm{(abs)}_r/(\alpha m^4)$, and $J^\mathrm{(scat)}_r/(\alpha m^4)$. The charge parameter $q = 1/2$; $z = 1$. Vertical lines mark the locations of the black-hole horizon and the photon sphere.}
    \label{fig:jr}
\end{figure}

\begin{figure}
    \centering
    \includegraphics[width=0.45\textwidth]{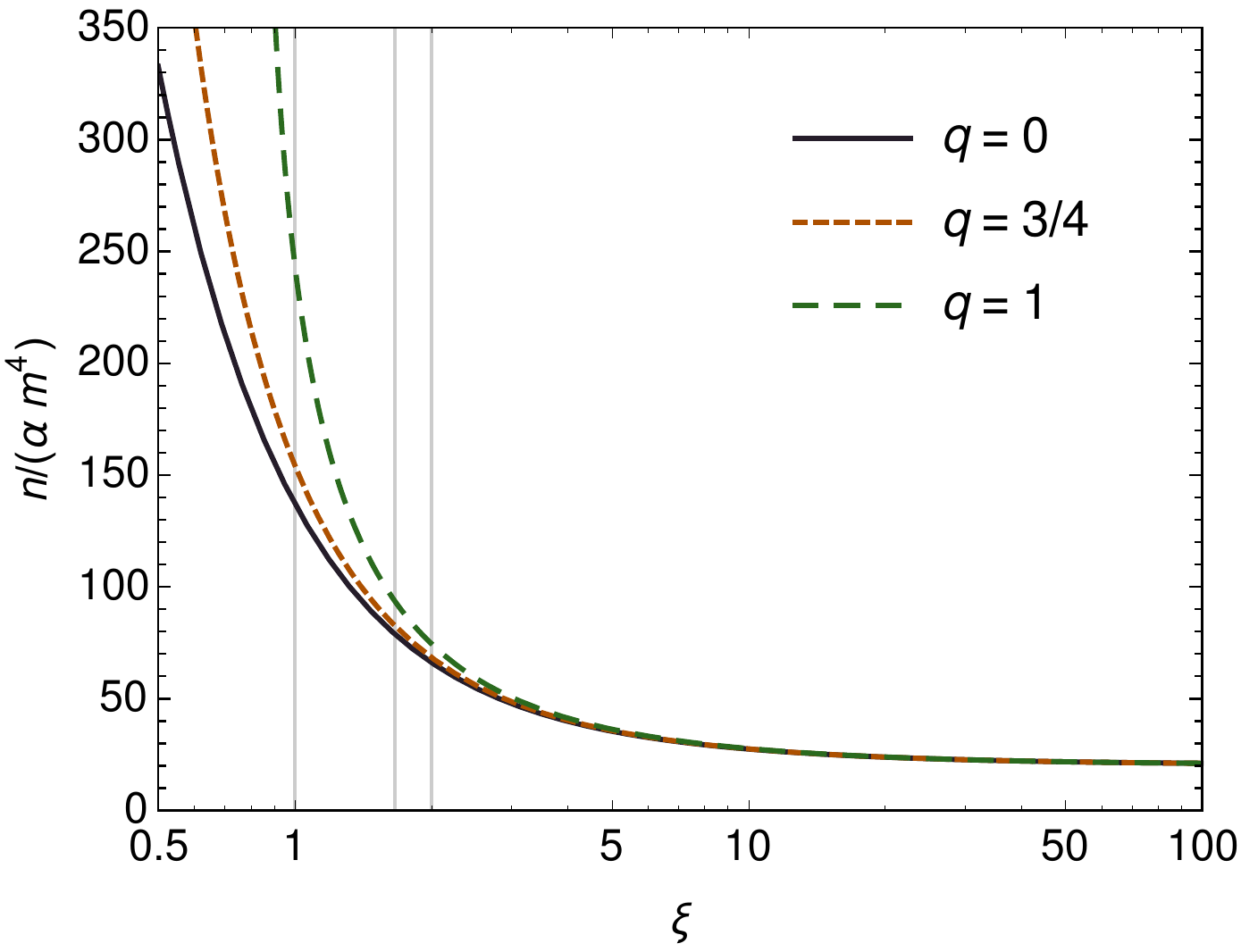}
    \caption{The particle density $n$ vs.\ $\xi$ for three Reissner-Nordstr\"{o}m solutions with $q = 0$ (Schwarzschild metric), $q = 3/4$, and $q = 1$ (extremal Reissner-Nordstr\"{o}m solution). All solutions are obtained assumming $z = 1$. Vertical lines mark the locations of the black-hole horizons for $q = 0$, $3/4$, and $1$.}
    \label{fig:n}
\end{figure}

\begin{figure}
    \centering
    \includegraphics[width=0.45\textwidth]{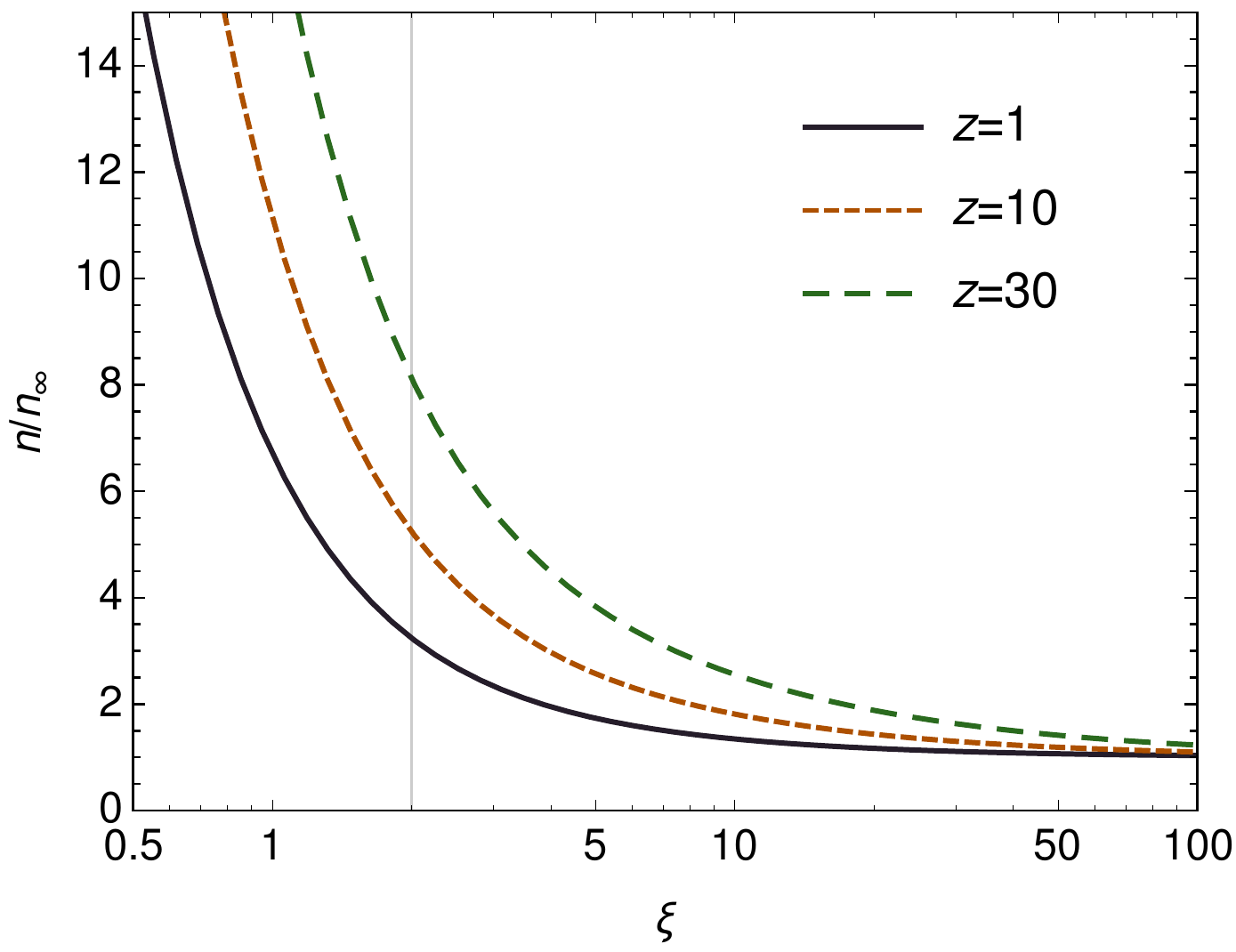}
    \caption{The ratio $n/n_\infty$ vs.\ $\xi$ for the Schwarzschild metric ($q = 0$). Different graphs correspond to $z = 1$, $10$, and $30$. The vertical line at $\xi = 2$ marks the location of the black-hole horizon.}
    \label{fig:n0}
\end{figure}

\begin{figure}
    \centering
    \includegraphics[width=0.45\textwidth]{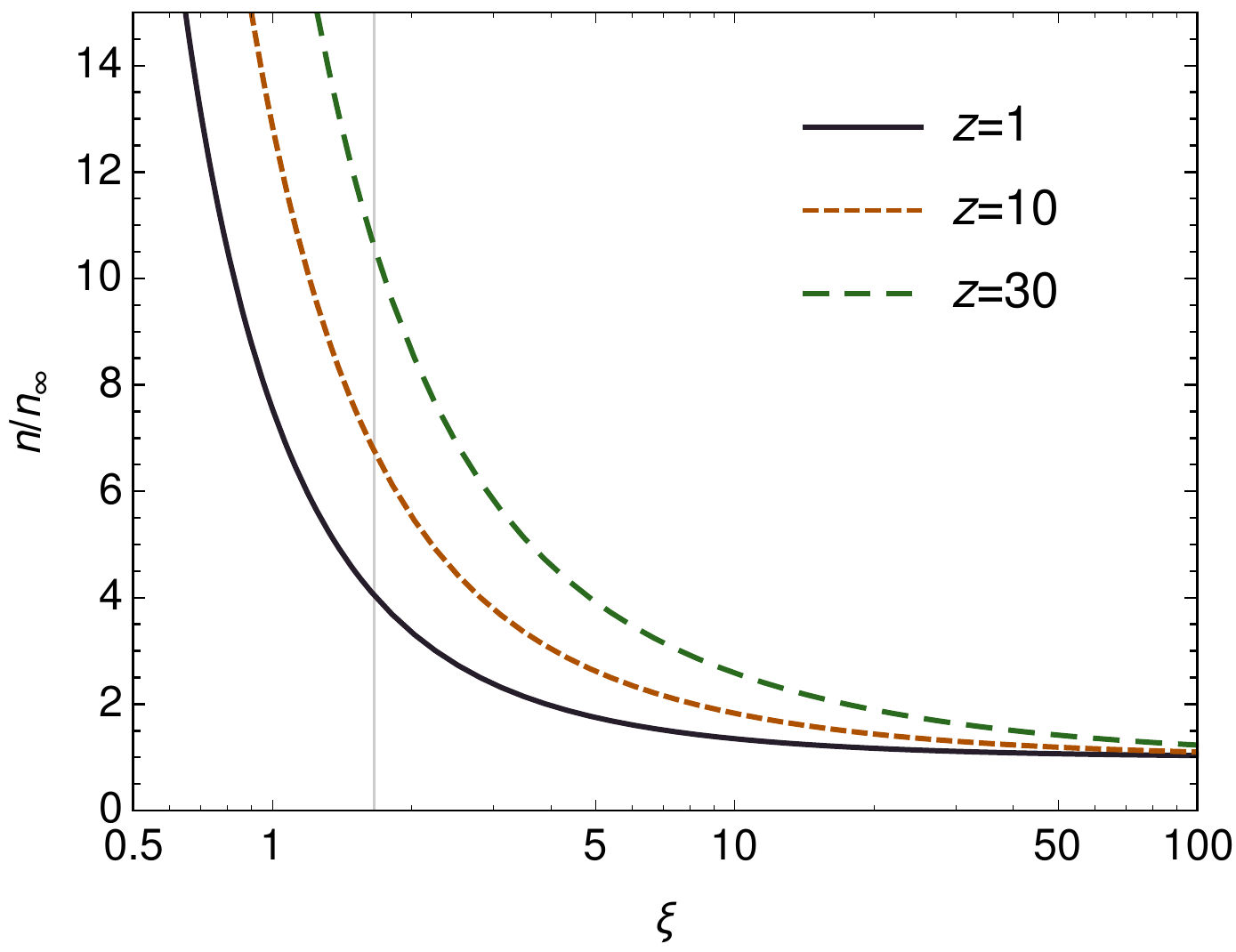}
    \caption{Same as in Fig.\ \ref{fig:n0} but for the Reissner-Nordstr\"{o}m solution with $q = 3/4$. The vertical line marks the location of the black-hole horizon.}
    \label{fig:n34}
\end{figure}

\begin{figure}
    \centering
    \includegraphics[width=0.45\textwidth]{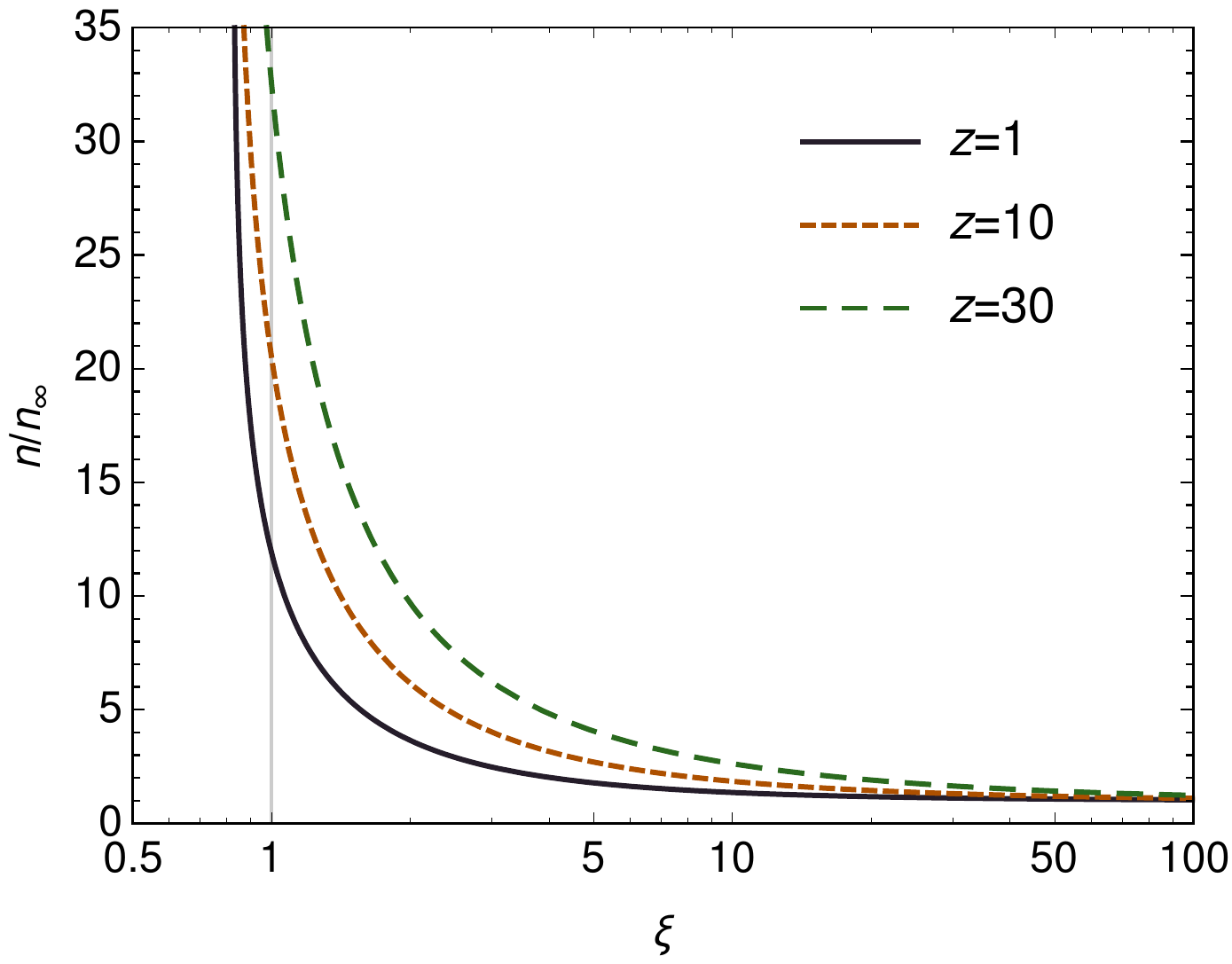}
    \caption{Same as in Fig.\ \ref{fig:n0} but for the extremal Reissner-Nordstr\"{o}m solution with $q = 1$. The vertical line at $\xi = 1$ marks the location of the black-hole horizon.}
    \label{fig:n1}
\end{figure}

\begin{figure}
    \centering
    \includegraphics[width=0.45\textwidth]{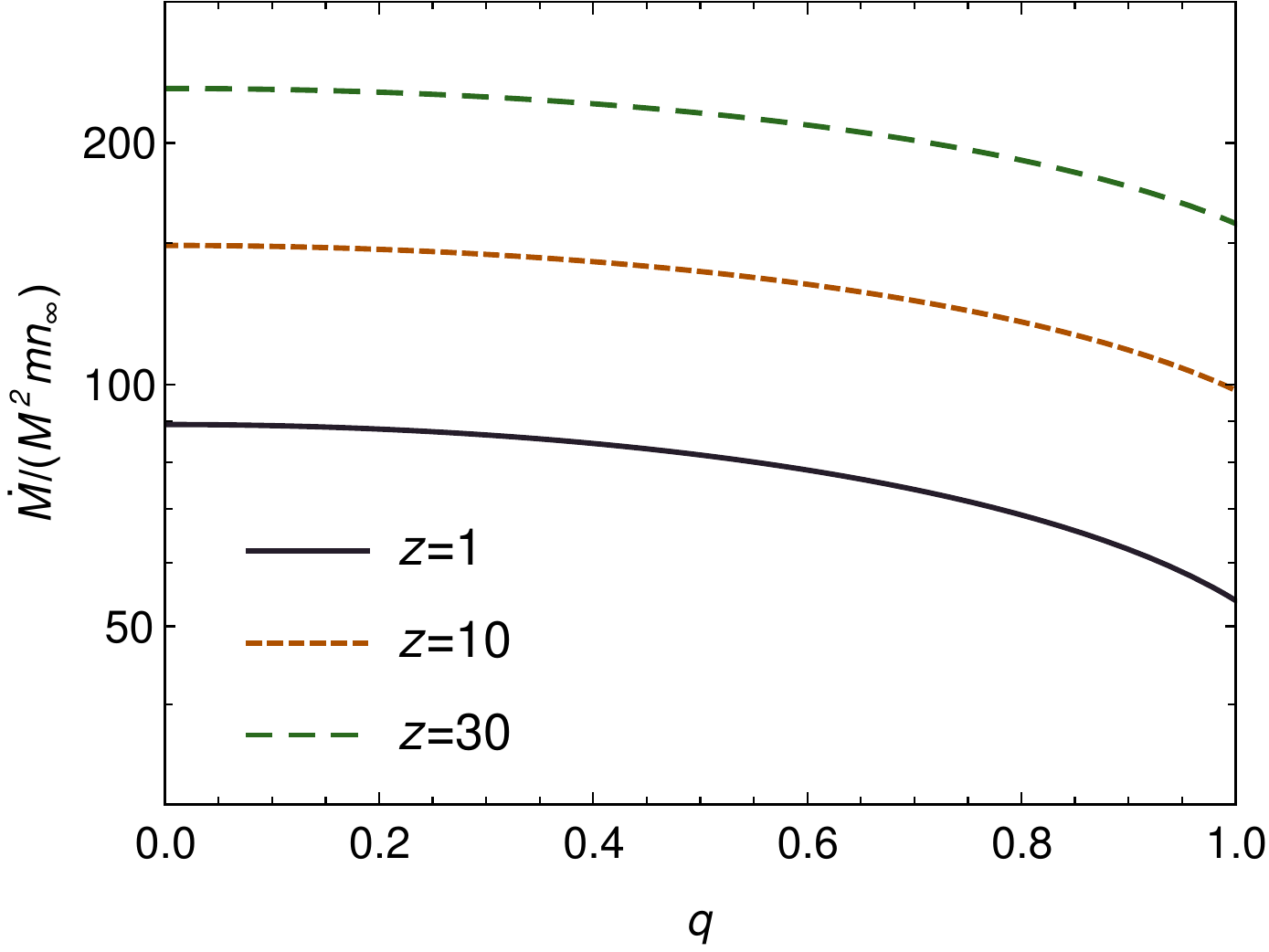}
    \caption{Mass accretion rate $\dot M/(M^2 m n_\infty)$ vs.\ the charge parameter $q$.}
    \label{fig:mdot1}
\end{figure}

\begin{figure}
    \centering
    \includegraphics[width=0.45\textwidth]{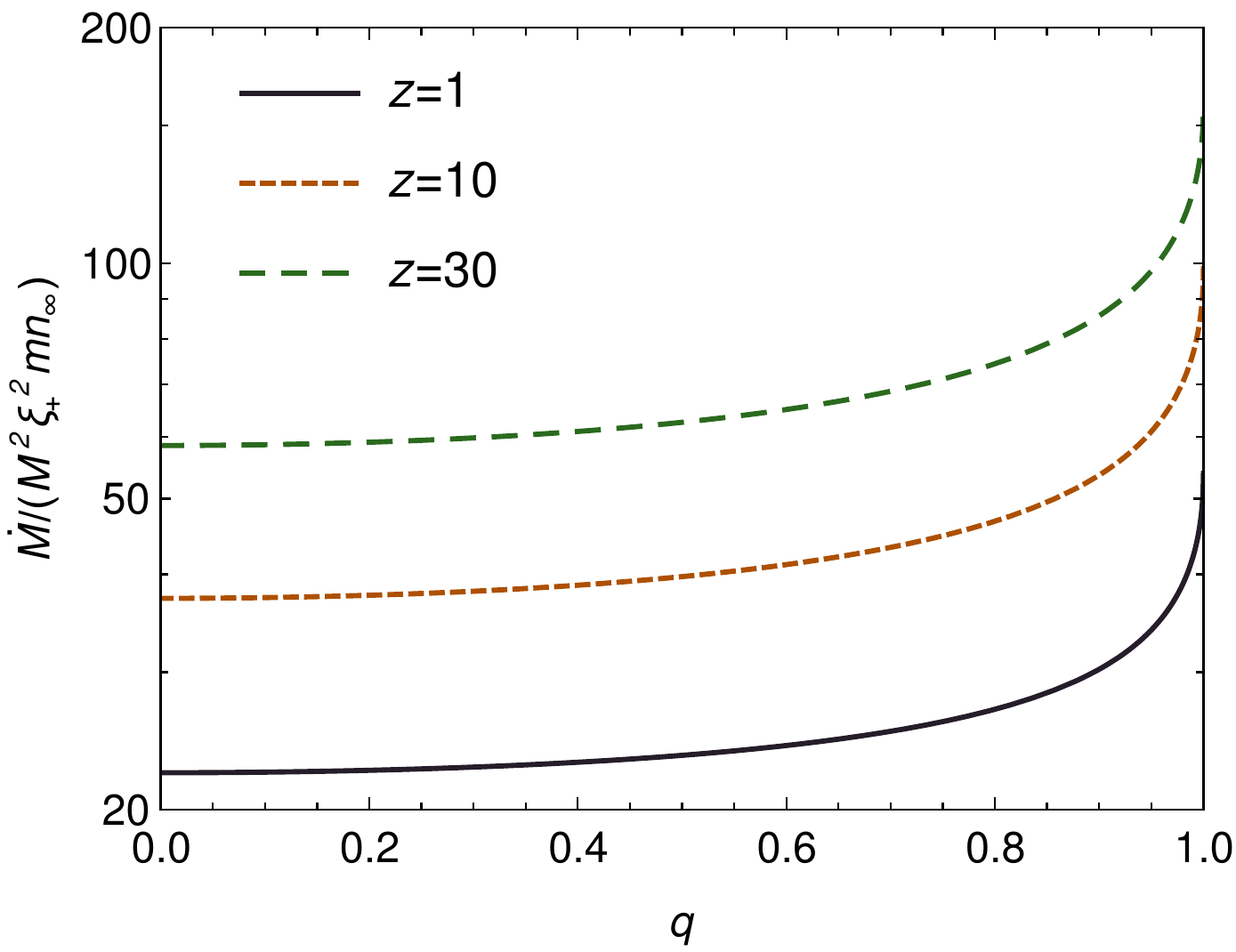}
    \caption{Mass accretion rate $\dot M/(M^2 \xi_+^2 m n_\infty)$ vs.\ the charge parameter $q$.}
    \label{fig:mdot2}
\end{figure}

The integrals derived in the preceding sections can be computed numerically with relatively little effort. We start illustrating our results with the graphs of the particle current density $J_\mu$. We work with standard Eddington--Finkelstein-type coordinates, i.e., we set $\eta \equiv 1$. Some generic plots of the components $J_t$ and $J_r$ are shown in Figs.\ \ref{fig:jt} and \ref{fig:jr} for $q = 1/2$ and $z = 1$. We show explicitly the two components $J^\mathrm{(scat)}_\mu$ and $J^\mathrm{(abs)}_\mu$, as well as the sum $J_\mu = J^\mathrm{(scat)}_\mu + J^\mathrm{(abs)}_\mu$. Note that although the total current $J_\mu$ is smooth, the components $J^\mathrm{(scat)}_\mu$ and $J^\mathrm{(abs)}_\mu$ are not. The components corresponding to the scattered particles vanish for $\xi < \frac{1}{2}\left( 3 + \sqrt{9 - 8 q^2} \right)$, i.e., below the photon sphere.

The particle density can be defined covariantly as
\[ n = \sqrt{-J_\mu J^\mu}. \]
It is a direct counterpart of the standard definition for perfect fluids, which relates the conserved particle current density with the four-velocity, i.e., $J^\mu = n u^\mu$. The product $mn$, which will appear frequently in this paper, is usually referred to as the rest-mass density.

Sample graphs of the particle density $n$ obtained for different black-hole charge parameters $q$ and for $z = 1$ are shown in Fig.\ \ref{fig:n}. Clearly, $n$ grows with the increasing charge parameter.

In all Figs.\ \ref{fig:jt}--\ref{fig:n} the locations of the black-hole horizons are marked with vertical lines. The areal radius of the black hole decreases from $r = 2M$ (or $\xi = 2$) for $q = 0$ to $r = M$ (or $\xi = 1$) for $q = 1$. Since $n$ is a decreasing function of the radius $\xi$, the increase of $n$ measured at the black hole horizon for the increasing charge parameter $q$ is even more pronounced.

The graphs of the particle density in Fig.\ \ref{fig:n} were normalized by $\alpha m^4$. It is a natural normalization, as long as we restrict ourselves to solutions corresponding to the same asymptotic temperature. To compare solutions with different $z$ we follow \cite{Olivier2} and normalize $n$ by its asymptotic value $n_\infty$. Sample plots of $n/n_\infty$ for $q = 0$, $3/4$, $1$ and $z = 1$, $10$, $30$ are shown in Figs.\ \ref{fig:n0}--\ref{fig:n1}. Figure \ref{fig:n0} shows the solutions obtained for $q=0$, i.e., assuming the Schwarzschild metric; it agrees with an analogous Fig.\ 2 in \cite{Olivier2}.

Another immediate result is the dependence of the mass accretion rate $\dot M$ given by Eq.\ (\ref{mdotcomputed}) on the black hole charge parameter $q$. A subtle point in comparing the accretion rates of different solutions is the proper choice of the normalization. In Fig.\ \ref{fig:mdot1} we plot $\dot M/(M^2 m n_\infty)$ vs.\ the charge parameter $q$ for three values of the asymptotic temperature $z = 1$, $10$, and $30$. In all cases the quantity $\dot M/(M^2 m n_\infty)$ decreases with the increasing $q$. Figure \ref{fig:mdot2} shows the dependence of $\dot M/(M^2 \xi_+^2 m n_\infty)$ on $q$. The normalization of $\dot M$ by $M^2 \xi_+^2 m n_\infty$ seems to be natural, as $4 \pi M^2 \xi_+^2$ is the area of the horizon. This normalization is also used in \cite{Olivier2}. On the other hand, since $\xi_+$ is a decreasing function of $q$, it changes the conclusion---$\dot M/(M^2 \xi_+^2 m n_\infty)$ increases with $q$. For Reissner-Nordstr\"{o}m black holes, for which the area of the horizon is not only a function of mass but also a charge parameter, we would rather opt for the first normalization.

The same behavior can be also observed for perfect fluids. In Appendix \ref{appendix:perfect_fluid} we give a short overview of a simple model of Bondi-type accretion of the perfect fluid on the Reissner-Nordstr\"{o}m black hole. It is obtained for the linear equation of state $p = k \rho$. Direct perfect-fluid equivalents of Figs.\ \ref{fig:mdot1} and \ref{fig:mdot2} are shown in Figs.\ \ref{fig:mdot3} and \ref{fig:mdot4}.

\subsection{Energy density, radial and tangential pressures}

\begin{figure}
    \centering
    \includegraphics[width=0.45\textwidth]{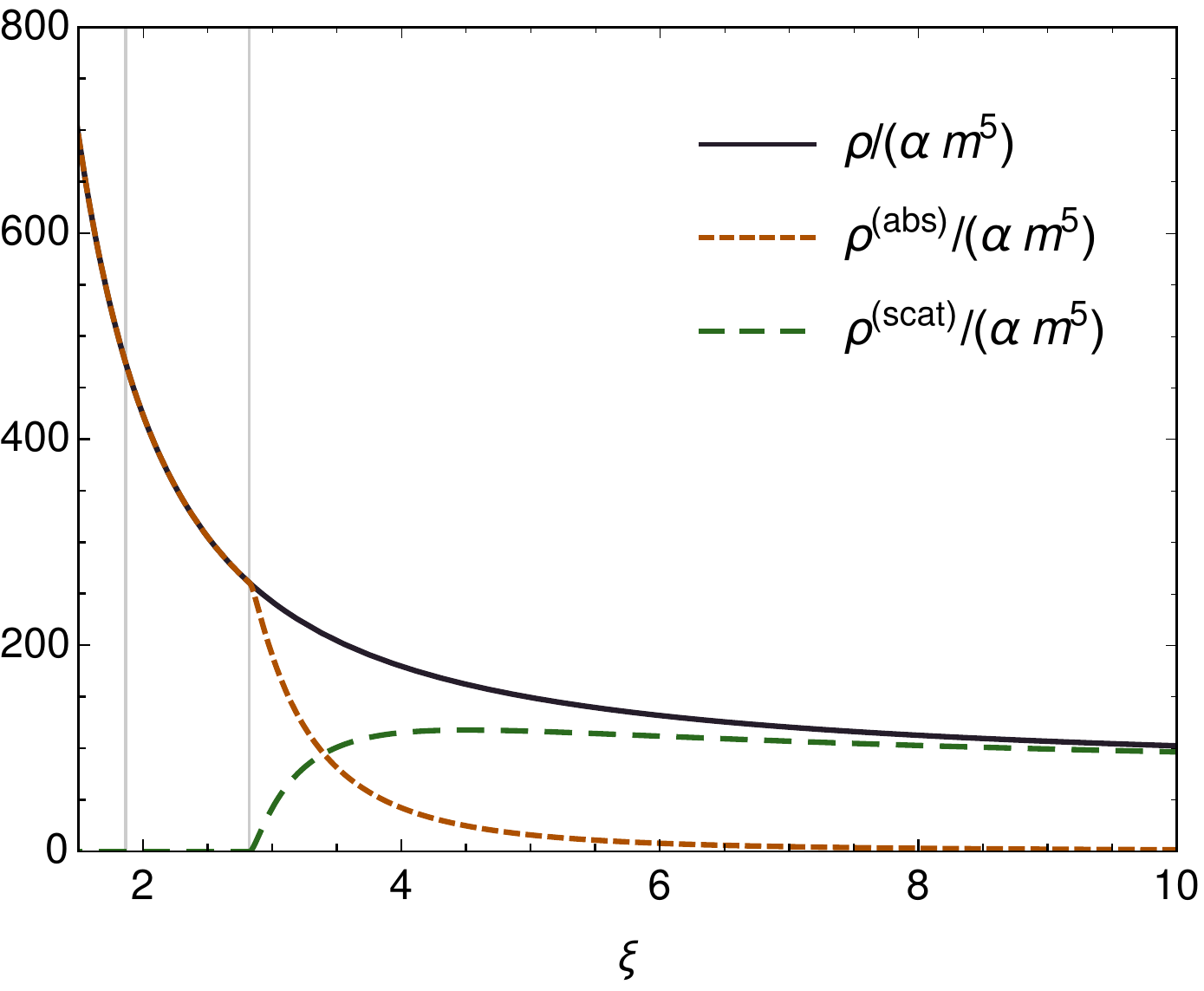}
    \caption{Sample graph of $\rho/(\alpha m^5)$, $\rho^\mathrm{(abs)}/(\alpha m^5)$, and $\rho^\mathrm{(scat)}/(\alpha m^5)$ for the parameters $q = 1/2$ and $z = 1$. Note that $\rho \neq \rho^\mathrm{(abs)} + \rho^\mathrm{(scat)}$, as explained in the text. The vertical lines mark the locations of the black-hole horizon and the photon sphere.}
    \label{fig:rhoq12}
\end{figure}

\begin{figure}
    \centering
    \includegraphics[width=0.45\textwidth]{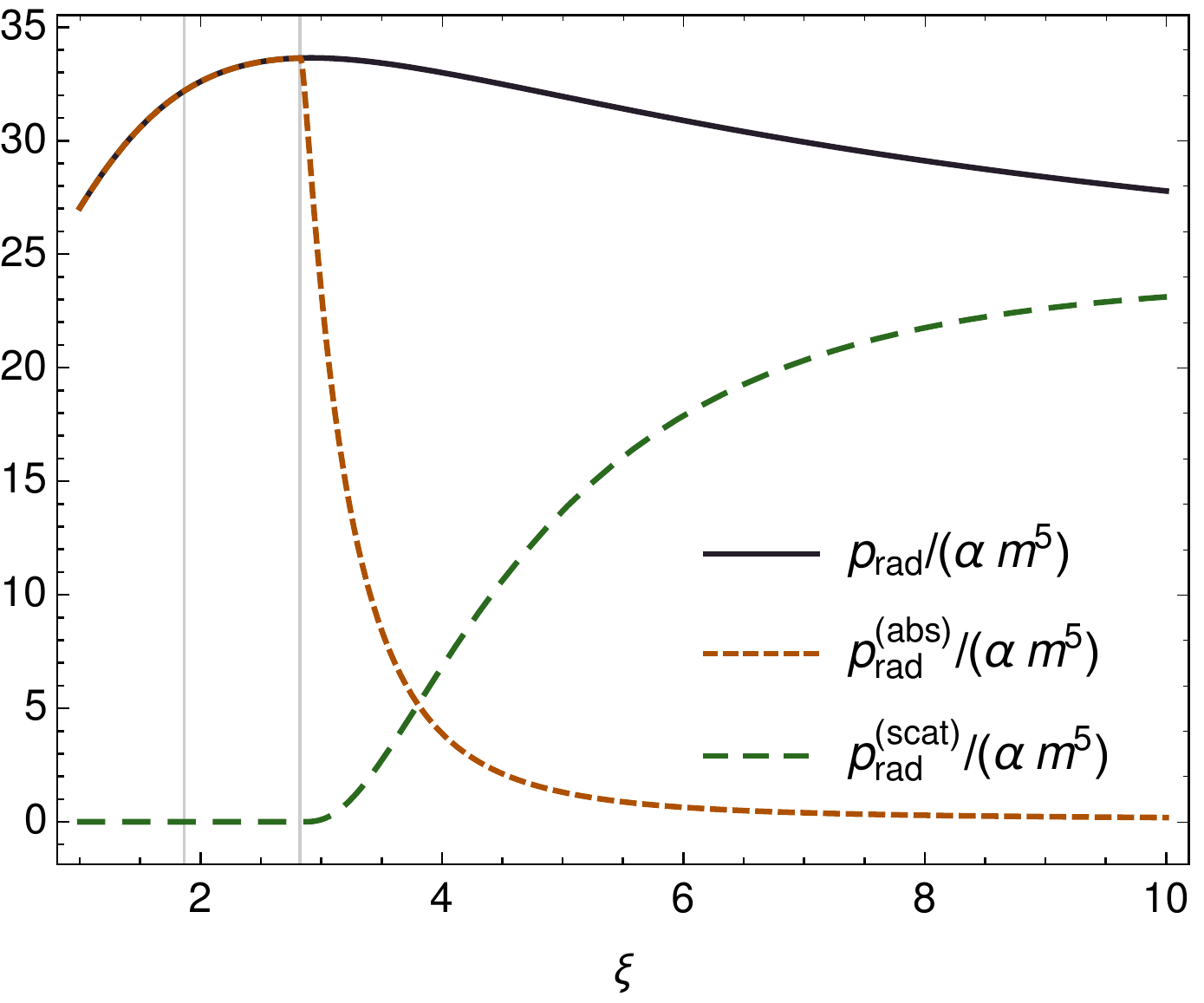}
    \caption{Sample graph of $p_\mathrm{rad}/(\alpha m^5)$, $p_\mathrm{rad}^\mathrm{(abs)}/(\alpha m^5)$, and $p_\mathrm{rad}^\mathrm{(scat)}/(\alpha m^5)$ for the parameters $q = 1/2$ and $z = 1$. Similarly to the energy densities, $p_\mathrm{rad} \neq p_\mathrm{rad}^\mathrm{(abs)} + p_\mathrm{rad}^\mathrm{(scat)}$, as explained in the text. The vertical lines mark the locations of the black-hole horizon and the photon sphere. Note that the radial pressure decreases in the vicinity of the horizon.}
    \label{fig:pradq12}
\end{figure}

\begin{figure}
    \centering
    \includegraphics[width=0.45\textwidth]{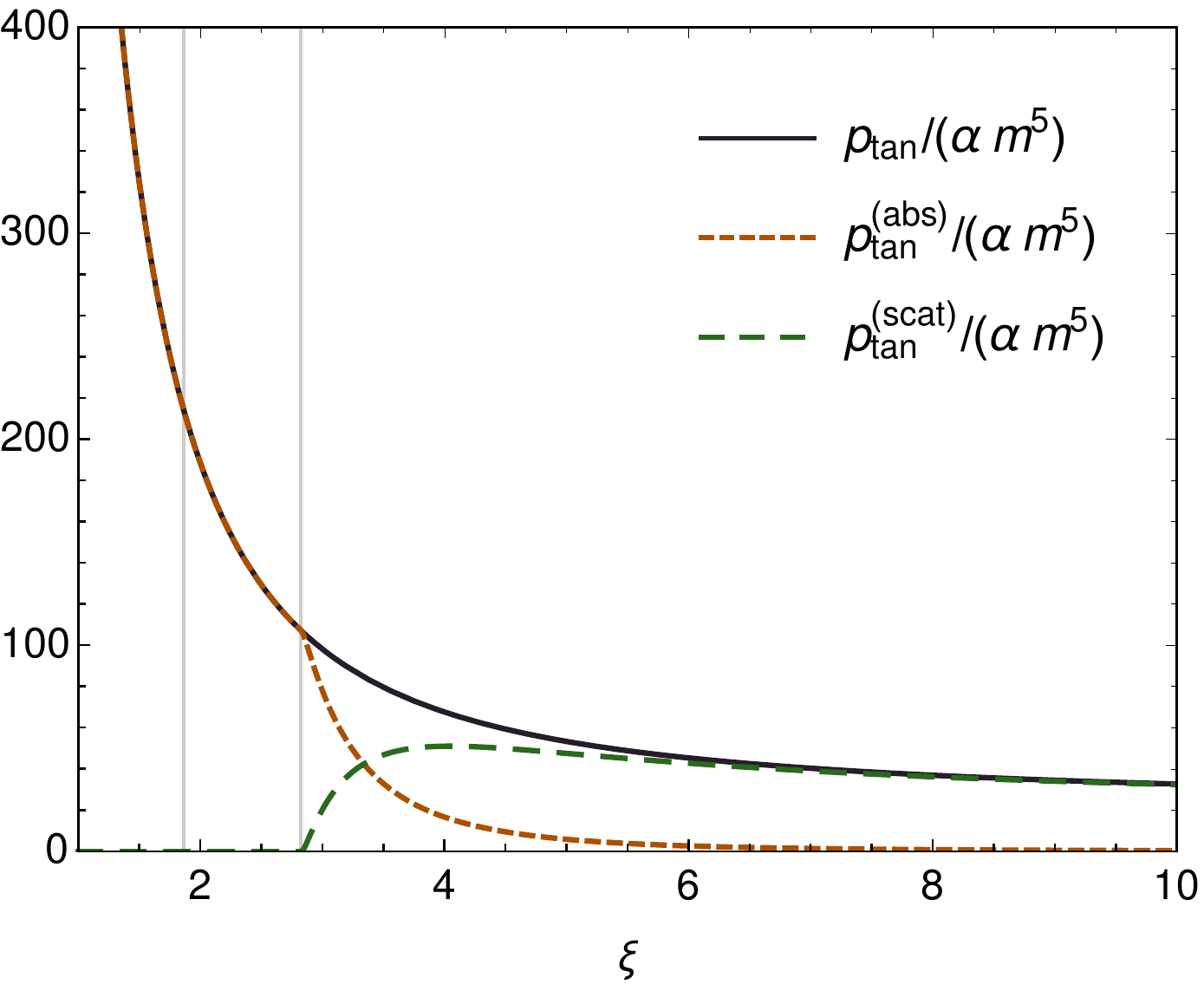}
    \caption{Sample graph of $p_\mathrm{tan}/(\alpha m^5)$, $p_\mathrm{tan}^\mathrm{(abs)}/(\alpha m^5)$, and $p_\mathrm{tan}^\mathrm{(scat)}/(\alpha m^5)$ for the parameters $q = 1/2$ and $z = 1$. The vertical lines mark the locations of the black-hole horizon and the photon sphere.}
    \label{fig:ptanq12}
\end{figure}

\begin{figure}
    \centering
    \includegraphics[width=\columnwidth]{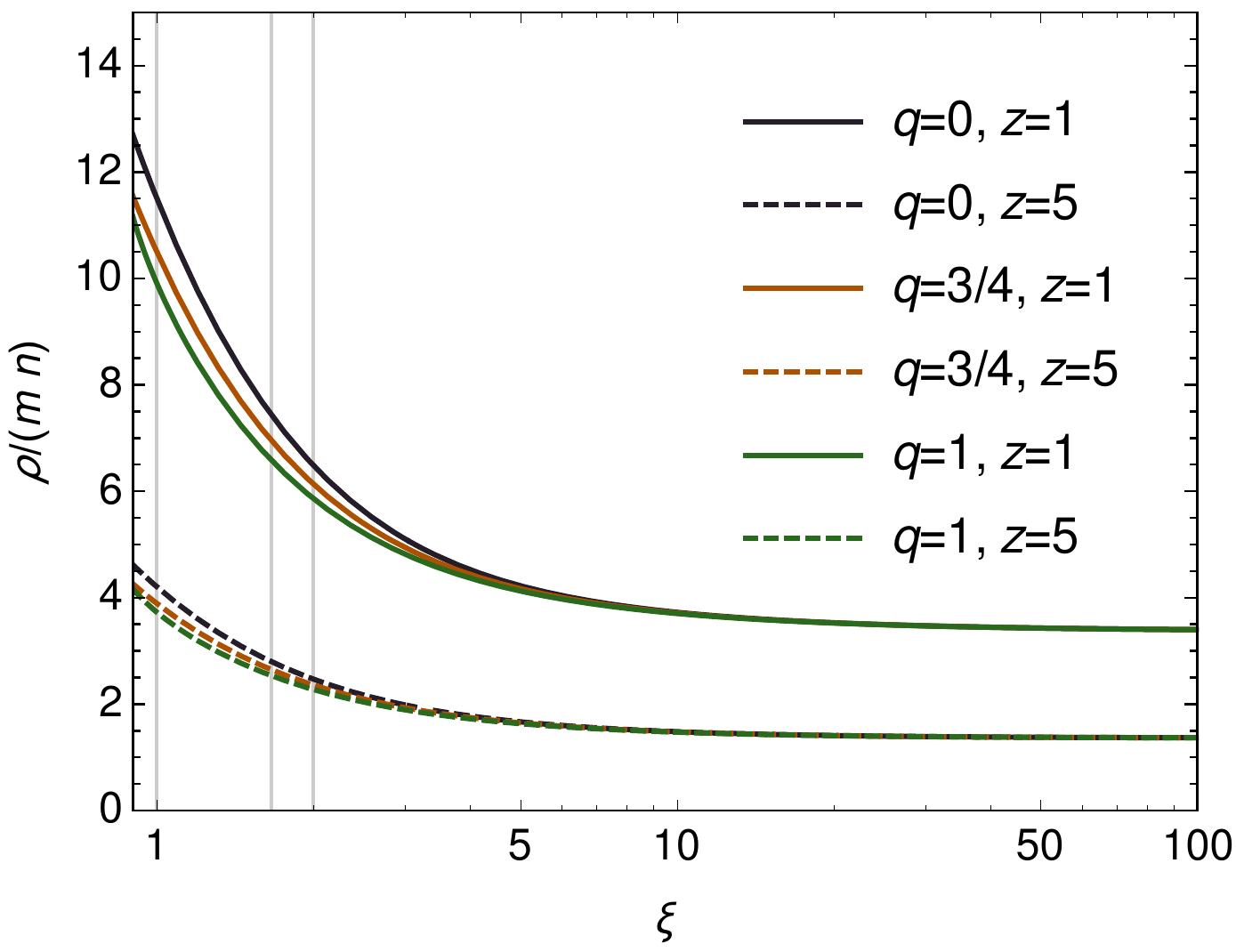}
    \caption{Sample graphs of the ratio $\rho/(m n)$ for $q = 0$, $3/4$, $1$, and $z = 1$, $5$. Vertical lines mark the locations of black-hole horizons for $q = 0$, $3/4$, and $1$.}
    \label{fig:rhon}
\end{figure}

\begin{figure}
    \centering
    \includegraphics[width=\columnwidth]{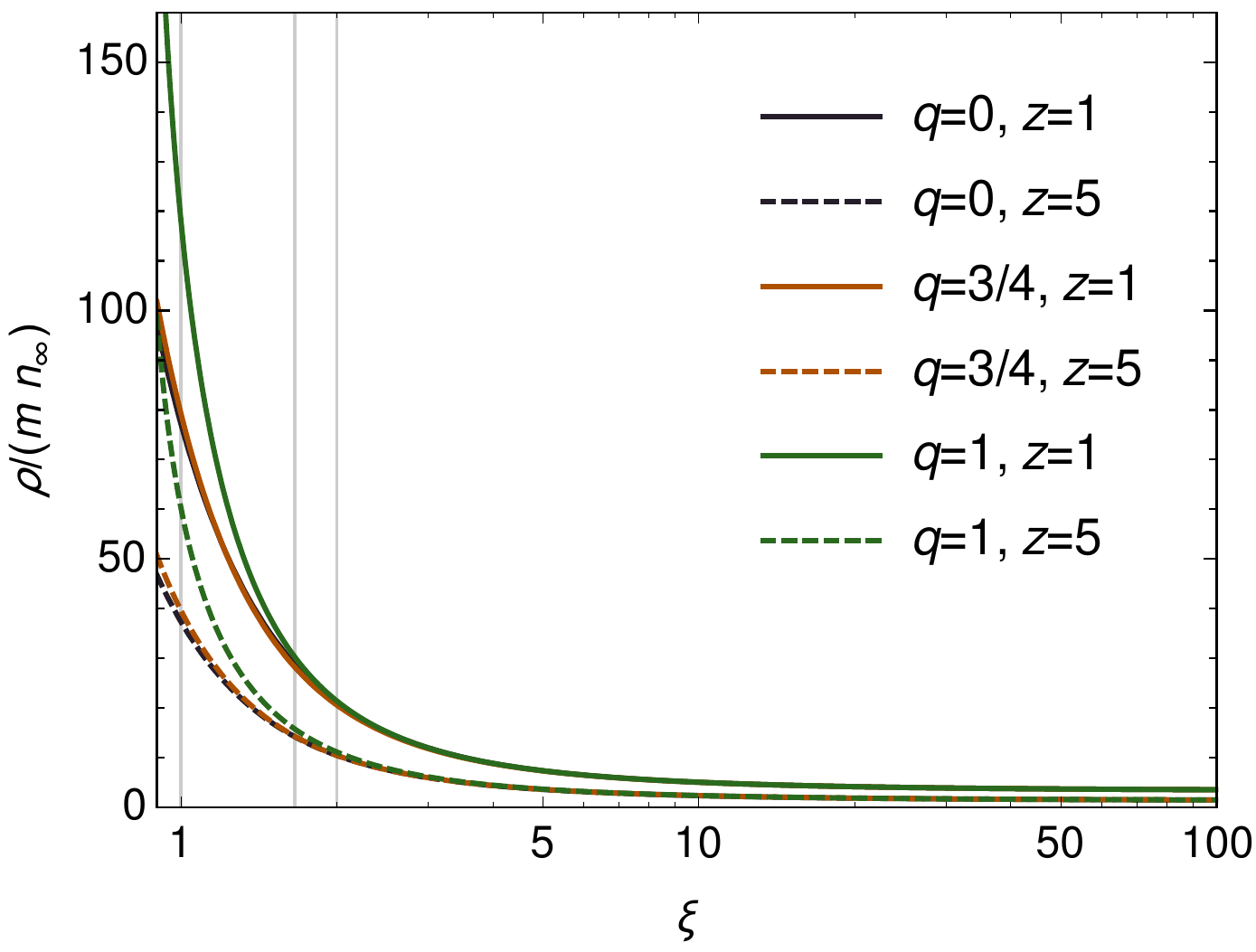}
    \caption{Sample graphs of the energy density $\rho$ normalized by $m n_\infty$ for $q = 0$, $3/4$, $1$, and $z = 1$, $5$. Vertical lines mark the locations of black-hole horizons for $q = 0$, $3/4$, and $1$.}
    \label{fig:rhoninf}
\end{figure}

\begin{figure}
    \centering
    \includegraphics[width=0.45\textwidth]{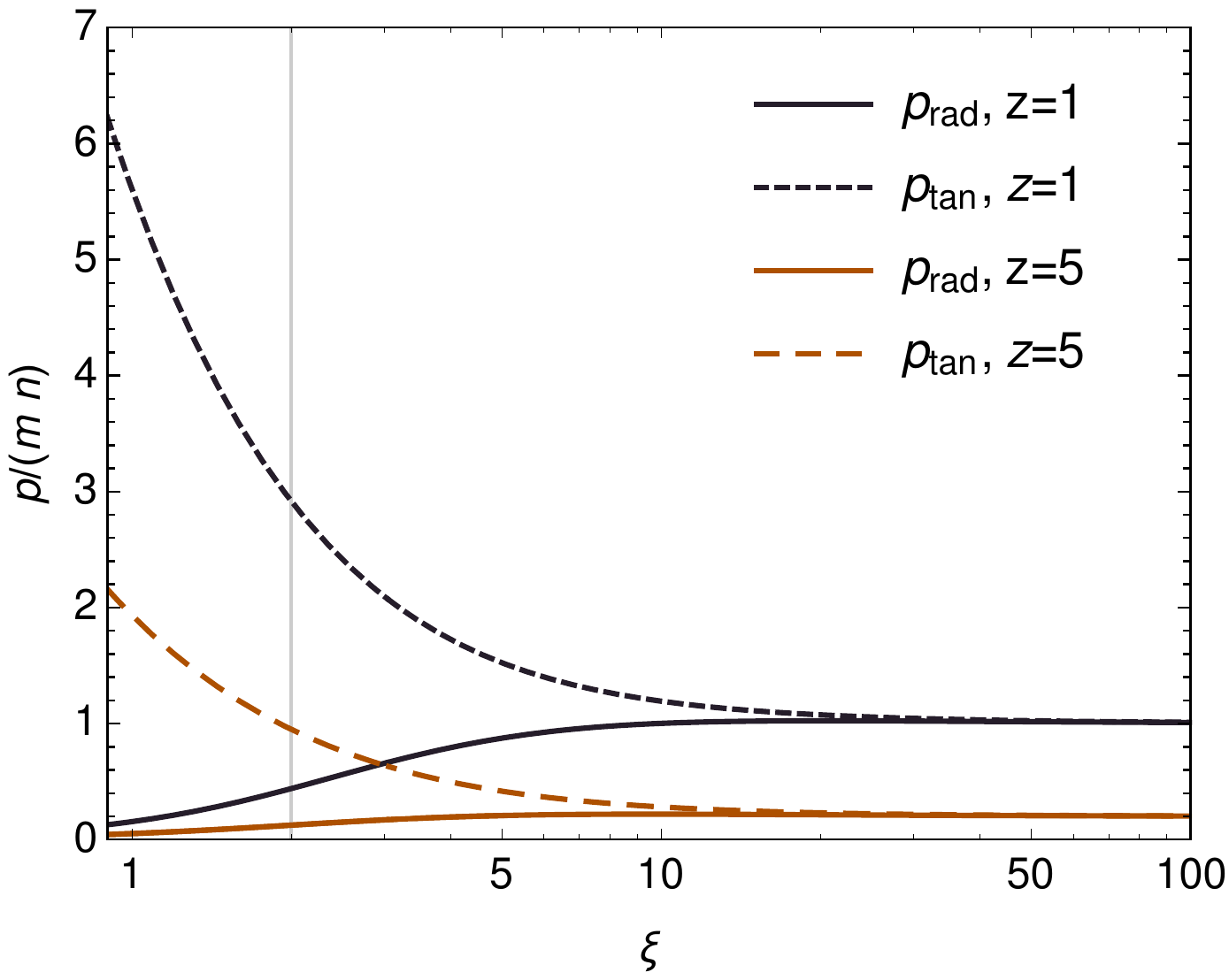}
    \caption{Sample graphs of the ratios $p_\mathrm{rad}/(mn)$ and $p_\mathrm{tan}/(mn)$ for $z = 1$, $5$, and $q = 0$ (Schwarzschild metric). The vertical line at $\xi = 2$ marks the location of the black-hole horizon.}
    \label{fig:pressuresq0}
\end{figure}

\begin{figure}
    \centering
    \includegraphics[width=0.45\textwidth]{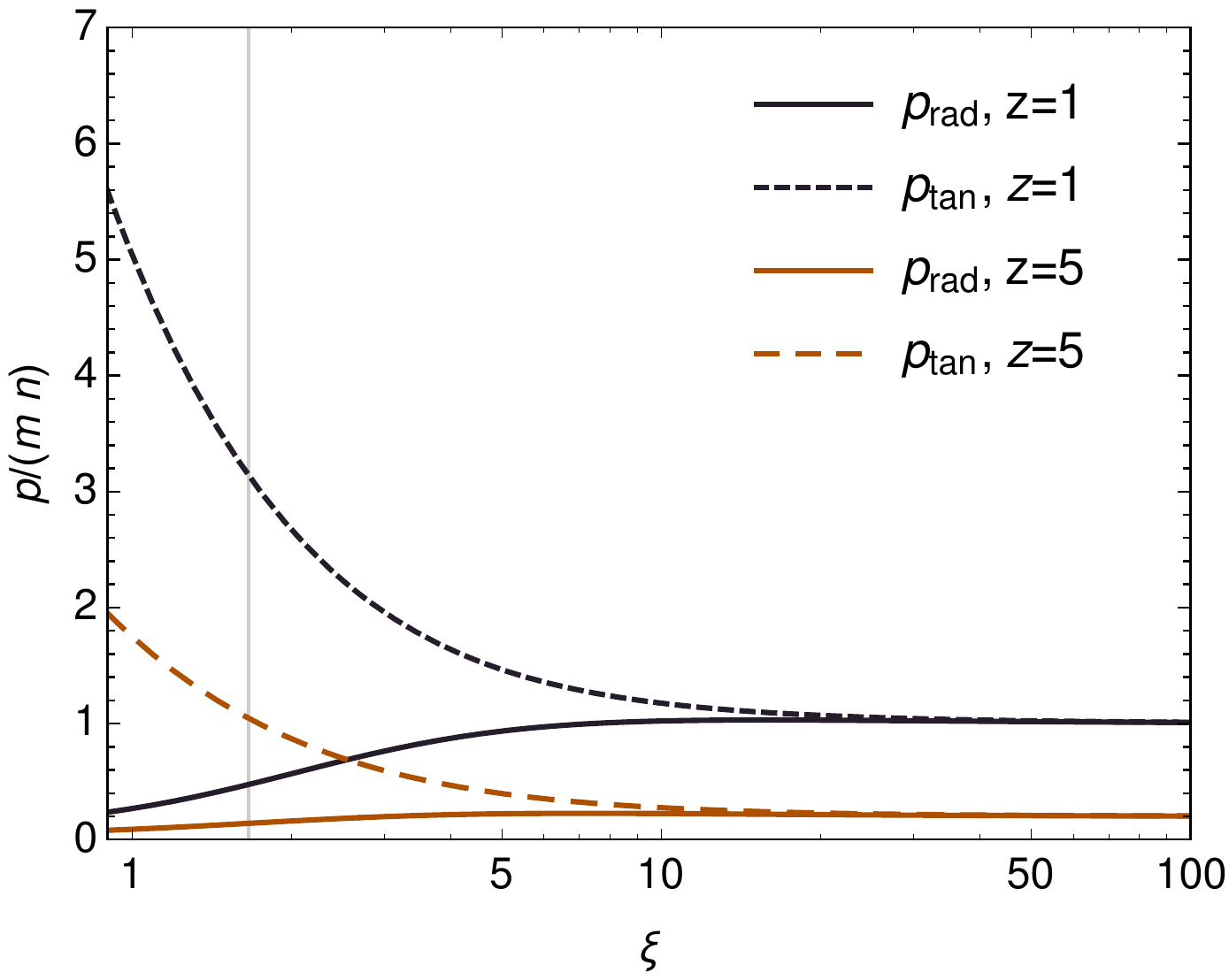}
    \caption{Sample graphs of the ratios $p_\mathrm{rad}/(mn)$ and $p_\mathrm{tan}/(mn)$ for $z = 1$, $5$, and $q = 3/4$. The vertical line marks the location of the black-hole horizon.}
    \label{fig:pressuresq34}
\end{figure}

\begin{figure}
    \centering
    \includegraphics[width=0.45\textwidth]{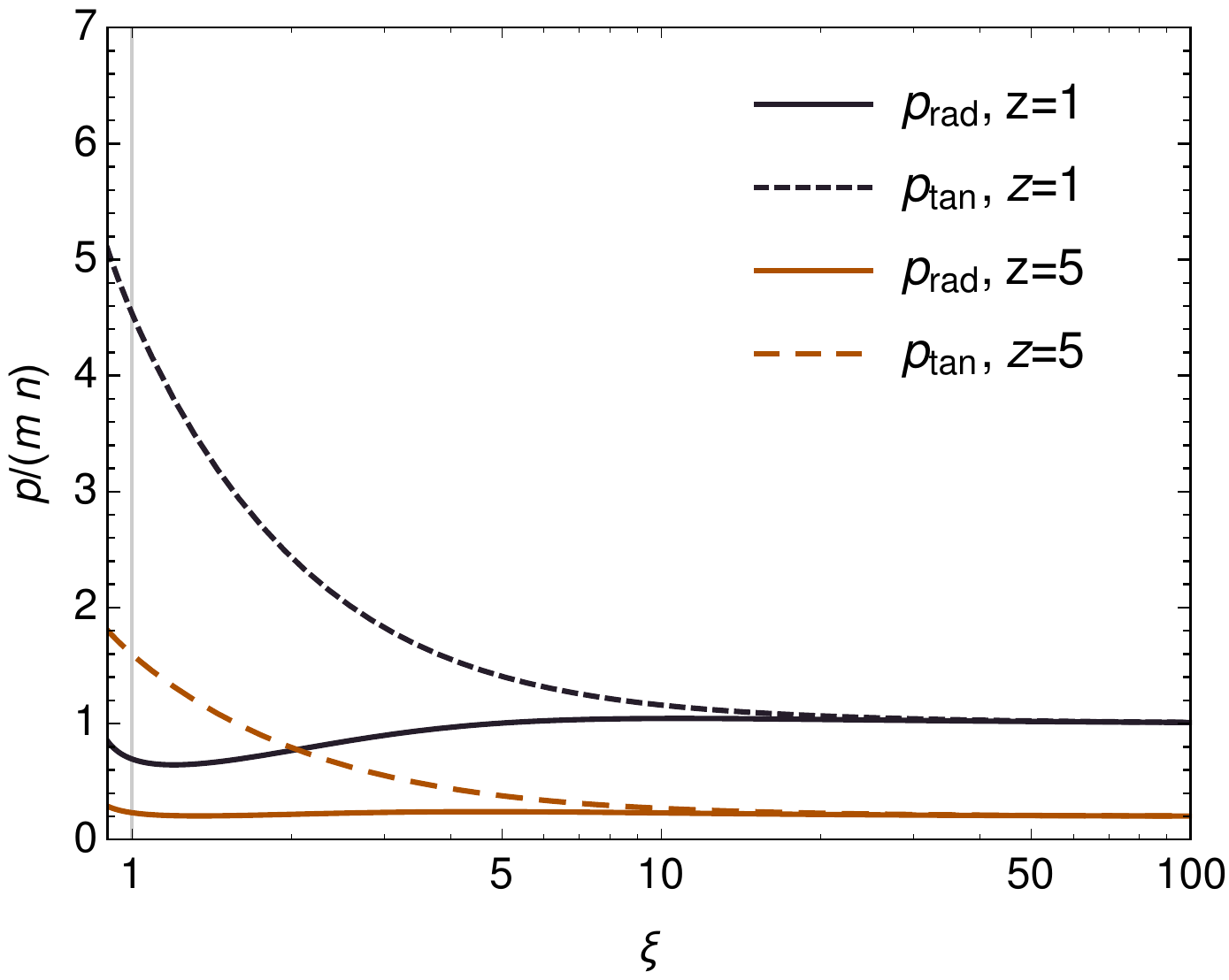}
    \caption{Sample graphs of the ratios $p_\mathrm{rad}/(mn)$ and $p_\mathrm{tan}/(mn)$ for $z = 1$, $5$, and $q = 1$ (extremal Reissner-Nordstr\"{o}m spacetime). The vertical line at $\xi = 1$ marks the location of the black-hole horizon.}
    \label{fig:pressuresq1}
\end{figure}

\begin{figure}
    \centering
    \includegraphics[width=\columnwidth]{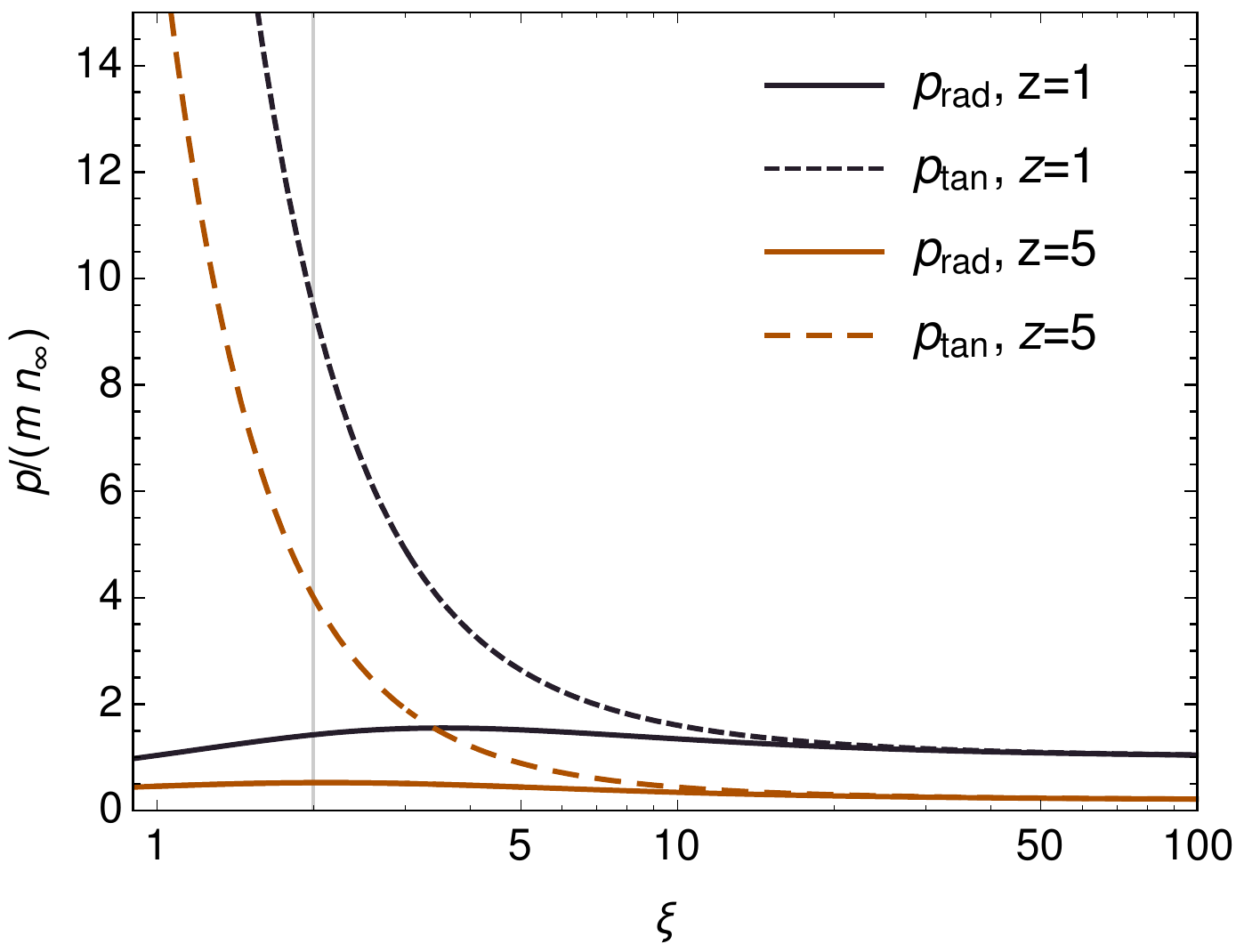}
    \caption{Sample graphs of radial and tangential pressures normalized by $m n_\infty$ for $z = 1$, $5$, and $q = 0$ (Schwarzschild metric). The vertical line at $\xi = 2$ marks the location of the black-hole horizon.}
    \label{fig:pressureinf0}
\end{figure}

\begin{figure}
    \centering
    \includegraphics[width=\columnwidth]{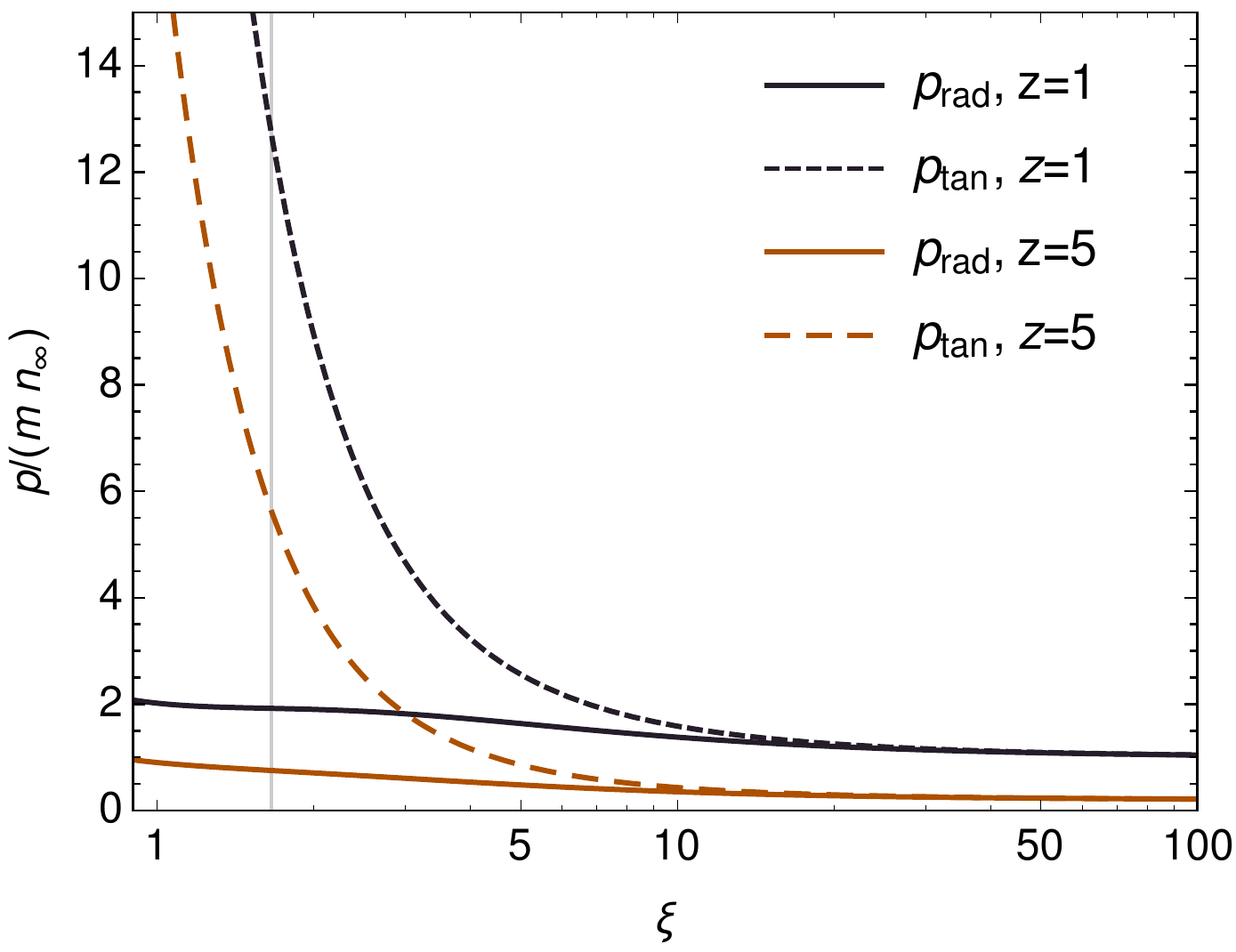}
    \caption{Sample graphs of radial and tangential pressures normalized by $m n_\infty$ for $z = 1$, $5$, and $q = 3/4$. The vertical line marks the location of the black-hole horizon.}
    \label{fig:pressureinf34}
\end{figure}

\begin{figure}
    \centering
    \includegraphics[width=\columnwidth]{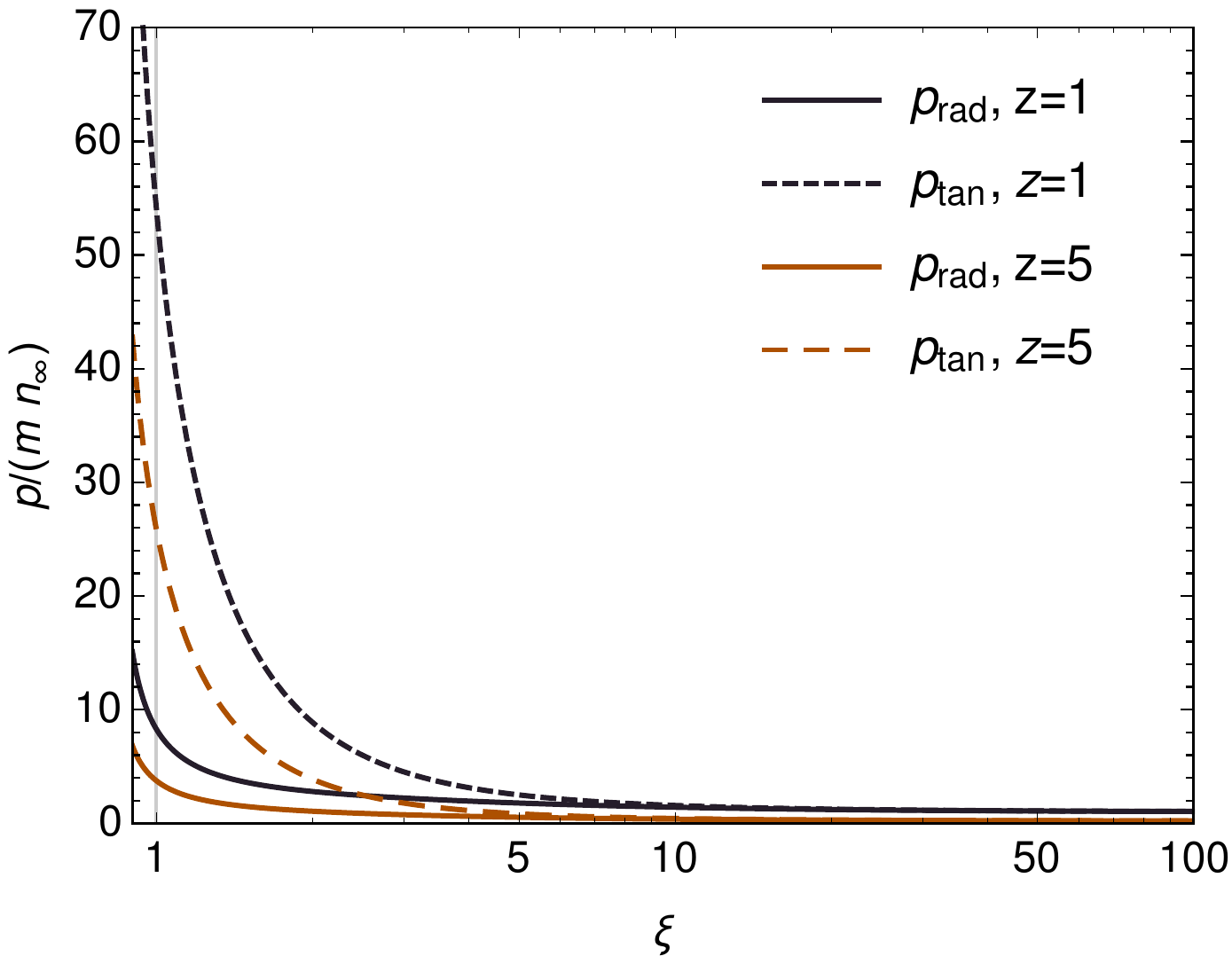}
    \caption{Sample graphs of radial and tangential pressures normalized by $m n_\infty$ for $z = 1$, $5$, and $q = 1$ (extremal Reissner-Nordstr\"{o}m spacetime). The vertical line at $\xi = 1$ marks the location of the black-hole horizon.}
    \label{fig:pressureinf1}
\end{figure}

\begin{figure}
    \centering
    \includegraphics[width=\columnwidth]{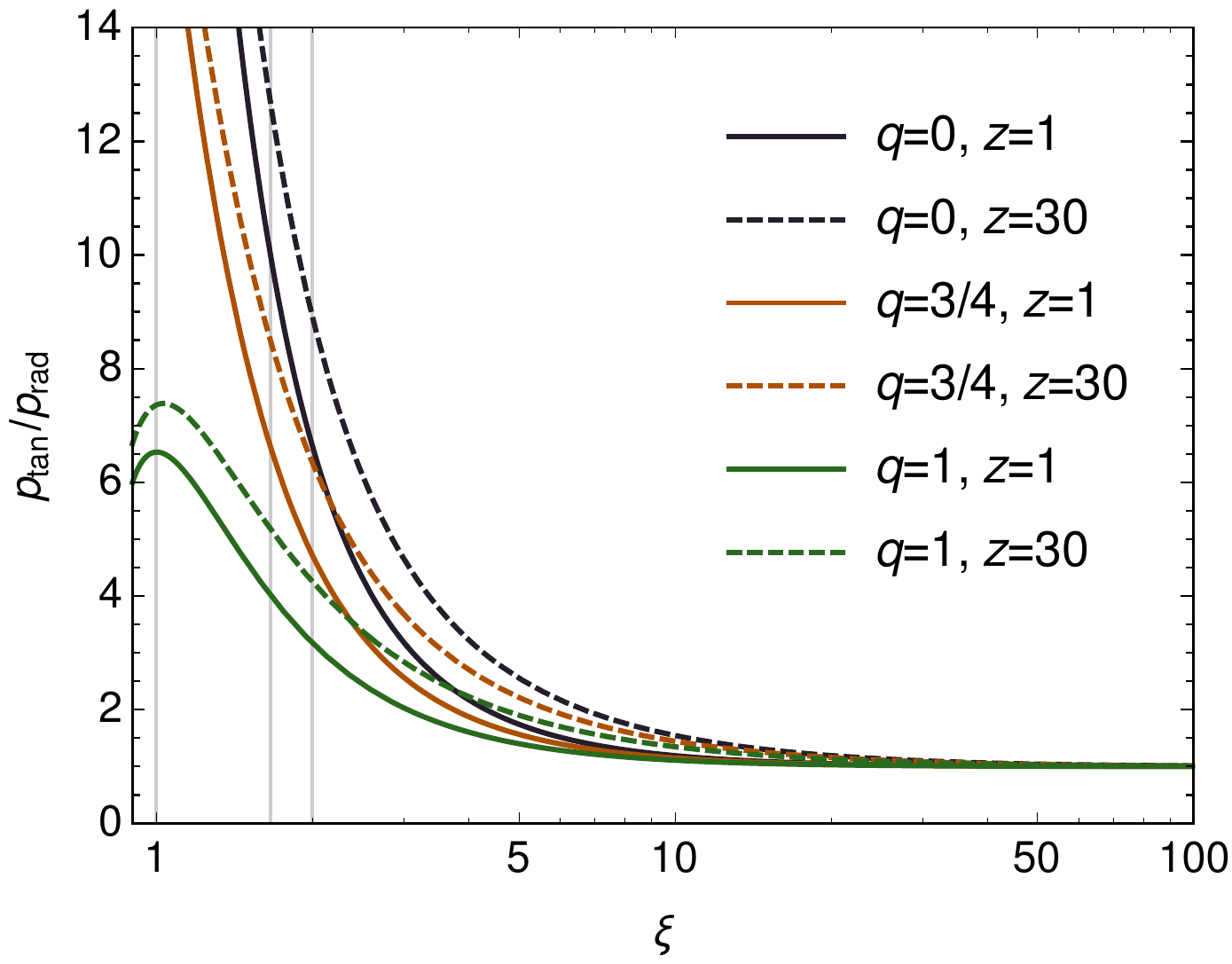}
    \caption{The ratio $p_\mathrm{tan}/p_\mathrm{rad}$ for $q = 0$, $3/4$ $1$, and $z = 1$, $30$. Vertical lines mark the locations of the black-hole horizon for $q = 0$, $3/4$, and $1$.}
    \label{fig:ptantoprad}
\end{figure}

Instead of plotting different components of $T\indices{^\mu_\nu}$, which similarily to $J_\mu$ are gauge-dependent, we concentrate on the eigenvalues $- \rho$, $p_\mathrm{rad}$, and $p_\mathrm{tan}$.

Sample plots of $\rho$, $\rho^\mathrm{(abs)}$, $\rho^\mathrm{(scat)}$, $p_\mathrm{rad}$, $p_\mathrm{rad}^\mathrm{(abs)}$, and $p_\mathrm{rad}^\mathrm{(scat)}$, $p_\mathrm{tan}$, $p_\mathrm{tan}^\mathrm{(abs)}$, and $p_\mathrm{tan}^\mathrm{(scat)}$ are shown in Figs.\ \ref{fig:rhoq12}--\ref{fig:ptanq12} for $q = 1/2$ and $z = 1$.  We normalize all these quantities by $\alpha m^5$. As discussed in Sec.\ \ref{sec:gasinequilibrium}, we have $p_\mathrm{tan} = p^\mathrm{(abs)}_\mathrm{tan} + p^\mathrm{(scat)}_\mathrm{tan}$, but $p_\mathrm{rad} \neq p^\mathrm{(abs)}_\mathrm{rad} + p^\mathrm{(scat)}_\mathrm{rad}$ and $\rho \neq \rho^\mathrm{(abs)} + \rho^\mathrm{(scat)}$. Also note that all terms associated with scattered particles vanish inside the photon sphere.

Figures \ref{fig:rhon} and \ref{fig:rhoninf} depict, respectively, the ratios $\rho/(m n)$ and $\rho/(m n_\infty)$ for $q = 0$, $3/4$, $1$, and $z = 1$, $5$. For both ratios, the dependence on the asymptotic temperature (i.e., on $z$) seems to be much stronger than the dependence on the charge parameter $q$.

Since asymptotically the gas is assumed to be in thermal equilibrium, both pressures $p_\mathrm{rad}$, and $p_\mathrm{tan}$ should tend at infinity to the same value, depending only on $z$, $\alpha$ and $m$. As the particles approach the black hole from infinity, both $p_\mathrm{tan}$ and $p_\mathrm{rad}$ increase initially. On the other hand, in the vicinity of the black hole $p_\mathrm{tan}$ still increases, but $p_\mathrm{rad}$ can decrease. It was observed in \cite{Olivier2} that for the Schwarzschild black hole the tangential pressure can be nearly an order of magnitude grater than the radial one. We illustrate this behavior in Figs.\ \ref{fig:pressuresq0}--\ref{fig:ptantoprad}. Figures \ref{fig:pressuresq0}--\ref{fig:pressuresq1} show the two pressures $p_\mathrm{rad}$ and $p_\mathrm{tan}$ normalized by the product of the particle  density $n$ and the particle mass $m$ (the rest-mass density) for $z = 1$ and $z = 5$. Figure \ref{fig:pressuresq0} was obtained for the Schwarzschild metric, and it agrees with Fig.\ 5 in \cite{Olivier2}. Figures  \ref{fig:pressuresq34} and \ref{fig:pressuresq1} were obtained for $q = 3/4$ and $q = 1$ (extremal Reissner-Nordstr\"{o}m spacetime), respectively. In all cases the ratios $p_\mathrm{rad}/(mn)$ and $p_\mathrm{tan}/(mn)$ decrease with $z$. Note, also that for $q = 1$ the ratio $p_\mathrm{rad}/(mn)$ can have both a local minimum and a local maximum outside the black-hole horizon. For comparison, we plot the same data in Figs.\ \ref{fig:pressureinf0}--\ref{fig:pressureinf1}, normalizing $p_\mathrm{tan}$ and $p_\mathrm{rad}$ by the asymptotic rest-mass density $m n_\infty$, instead of $mn$.

In Fig.\ \ref{fig:ptantoprad} we plot the ratio $p_\mathrm{tan}/p_\mathrm{rad}$ for $z = 1$, $30$, and $q = 0$, $3/4$, $1$. Vertical lines in this figure mark the locations of the black-hole horizons. The ratio $p_\mathrm{tan}/p_\mathrm{rad}$ grows with $z$, and it saturates relatively quickly---the graph of the ratio $p_\mathrm{tan}/p_\mathrm{rad}$ for $z = 30$ in Fig.\ \ref{fig:ptantoprad} would almost coincide with the graph obtained for, say, $z = 100$. On the other hand, the ratio $p_\mathrm{tan}/p_\mathrm{rad}$ generally decreases with the charge parameter $q$. Below we list a few sample values of $p_\mathrm{tan}/p_\mathrm{rad}$.
\begin{itemize}
    \item At the  black-hole horizons: 
        \begin{itemize}
            \item For $z=1$ and $q = 0$, $3/4$, $1$  we have $p_\mathrm{tan}/p_\mathrm{rad} = 6.65$, $6.62$, $6.53$, respectively
            \item For $z=30$ and $q =0$, $3/4$, $1$  we have $p_\mathrm{tan}/p_\mathrm{rad} = 8.93$, $8.49$, $7.36$, respectively
        \end{itemize}
    \item At the photon spheres:
        \begin{itemize}
            \item For $z = 1$ and $q = 0$, $3/4$, $1$  we have $p_\mathrm{tan}/p_\mathrm{rad} = 3.19$, $3.18$, $3.17$, respectively
            \item For $z=30$ and $q = 0$, $3/4$, $1$  we have $p_\mathrm{tan}/p_\mathrm{rad} = 4.66$, $4.49$, $4.26$, respectively
        \end{itemize}
\end{itemize}

\section{Conclusions}
\label{sec:conclusions}

Some features of the steady spherically-symmetric accretion of the Vlasov gas on Reissner-Nordstr\"{o}m black holes resemble those characteristic for the accretion of perfect fluids. Michel-type (or Bondi-type) accretion of perfect fluids on Reissner-Nordstr\"{o}m black holes was investigated, e.g., in 
\cite{babichev,ficek}; we give a short overview of this model in Appendix \ref{appendix:perfect_fluid}. As usual, precise results depend on the assumed equation of state. Several analytic solutions can be obtained for a class of linear equations of state of the form $p = k \rho$, where $0 < k \le 1$ is a constant. Similarly to the Vlasov case, for the black holes with a fixed mass, the mass accretion rate decreases with the increasing charge parameter. Also, as for the Vlasov model, the ratio of the particle density at the black-hole horizon to its asymptotic value (the so-called compression parameter) grows with the charge parameter $q$ (we show this fact in Appendix \ref{appendix:perfect_fluid} for the stiff equation of state with $k = 1$).

In terms of the energy-momentum tensor, the two models (the perfect fluid and the Vlasov model) differ significantly. In this work we have recovered the general properties of the energy-momentum tensor and its eigenvalues---the energy density, the radial and tangential pressures---discovered by Rioseco and Sarbach in \cite{Olivier} for the accretion of the Valsov gas on Schwarzschild black holes. Similarily to the Schwarzschild case, we observe that the two pressures differ in the vicinity of the black hole. While the tangential pressure is a decreasing function of the radius, the radial pressure is not monotonic, and it decreases near the black-hole horizon. In general, $p_\mathrm{tan}$ exceeds $p_\mathrm{rad}$ in the vicinity of the black hole, but the precise ratio $p_\mathrm{tan}/p_\mathrm{rad}$ depends both on the asymptotic temperature of the gas and on the black-hole charge parameter $q$. For $z = 1$ we have at the black-hole horizons $p_\mathrm{tan}/p_\mathrm{rad} = 6.65 $ for $q = 0$, and $p_\mathrm{tan}/p_\mathrm{rad} = 6.53$ for $q = 1$. The corresponding values for $z = 30$ are $p_\mathrm{tan}/p_\mathrm{rad} = 8.93 $ for $q = 0$, and $p_\mathrm{tan}/p_\mathrm{rad} = 7.36$ for $q = 1$.

The model presented in this paper follows the footsteps of Rioseco and Sarbach \cite{Olivier}, as closely, as possible. In particular, we have neglected a number of factors that could both complicate and alter the corresponding physical picture. In the first place, we have neglected all terms describing the scattering between the particles of the gas. As a consequence, there is no interaction between the two classes of particles investigated in this paper: those absorbed by the black hole, and those scattered to infinity. More importantly, neglecting the scattering between the particles, one can also neglect the existence of particles on bounded trajectories. The latter would also become important, if we took into account the self-gravity of the accreting gas and attempted to solve the corresponding Einstein-Vlasov system.

In the analysis presented in this paper we have deliberately concentrated only on stationary states. The formalism developed in \cite{Olivier} allows also for a relatively simple stability analysis, which we postpone for future.

From the perspective of the current work the most interesting direction of the future work is the analysis of axially symmetric accretion systems (we are mainly interested in the Reissner-Nordstr\"{o}m metric as a toy model for the spinning black hole). An elegant work in this direction has recently been published by Rioseco and Sarbach, who investigated the motion of Vlasov gas in the equatorial plane of the Kerr spacetime \cite{olivier_kerr}.

\begin{acknowledgments}
We would like to thank Olivier Sarbach for discussions and comments. PM was partially supported by the Polish National Science Centre grant No.\ 2017/26/A/ST2/00530.
\end{acknowledgments}

\appendix
\section{Integrals with respect to the angular momentum}

In this appendix we collect a few analytic integrals used in establishing the results of Sec.\ \ref{sec:gasinequilibrium}. For simplicity we denote
\[ s_1 =  \sqrt{\varepsilon^2 - U_{\lambda_1}(\xi)}, \quad s_2 =  \sqrt{\varepsilon^2 - U_{\lambda_2}(\xi)}.\]

\begin{eqnarray*}
\lefteqn{\int_{\lambda_1}^{\lambda_2} \frac{\lambda d\lambda}{\sqrt{\varepsilon^2 - U_\lambda(\xi)}} = \frac{\lambda_2^2 - \lambda_1^2}{s_1 + s_2}, }\\
\lefteqn{\int_{\lambda_1}^{\lambda_2} \frac{\lambda^3 d\lambda}{\sqrt{\varepsilon^2 - U_\lambda(\xi)}} =} \\
&& \frac{\lambda_2^2 - \lambda_1^2}{3 (s_1 + s_2)^2} \left[ \lambda_2^2 (2 s_1 + s_2) + \lambda_1^2 (s_1 + 2 s_2) \right], \\
\lefteqn{\int_{\lambda_1}^{\lambda_2} \lambda \sqrt{\varepsilon^2 - U_\lambda(\xi)} d \lambda = - \frac{\xi^2 (s_2 - s_1)}{3N},} \\
\lefteqn{\int_{\lambda_1}^{\lambda_2} \pi_{\xi \pm} \frac{\lambda d\lambda}{\sqrt{\varepsilon^2 - U_\lambda(\xi)}} = \frac{\lambda_2^2 - \lambda_1^2}{N} \left[ \frac{(1 - N \eta) \varepsilon}{s_1 + s_2} \pm \frac{1}{2} \right], }\\
 \lefteqn{\int_{\lambda_1}^{\lambda_2} \pi_{\xi \pm}^2 \frac{\lambda d\lambda}{\sqrt{\varepsilon^2 - U_\lambda(\xi)}} = }\\
 &&\frac{\lambda_2^2 - \lambda_1^2}{ N^2 (s_1 + s_2)} \biggl\{ ((1 - N \eta) \varepsilon)^2 + \varepsilon^2 - N  \\
 && \pm (1 - N \eta) \varepsilon (s_1 + s_2) \\
 &&  - \frac{N}{3 \xi^2 (s_1 + s_2)} \left[ \lambda_2^2 (2 s_1 + s_2) + \lambda_1^2 (s_1 + 2 s_2) \right]  \biggr\}.
\end{eqnarray*}

\section{Spherically symmetric accretion of perfect fluids in the Reissner-Nordstr\"{o}m spacetime}
\label{appendix:perfect_fluid}

\begin{figure}
    \centering
    \includegraphics[width=0.45\textwidth]{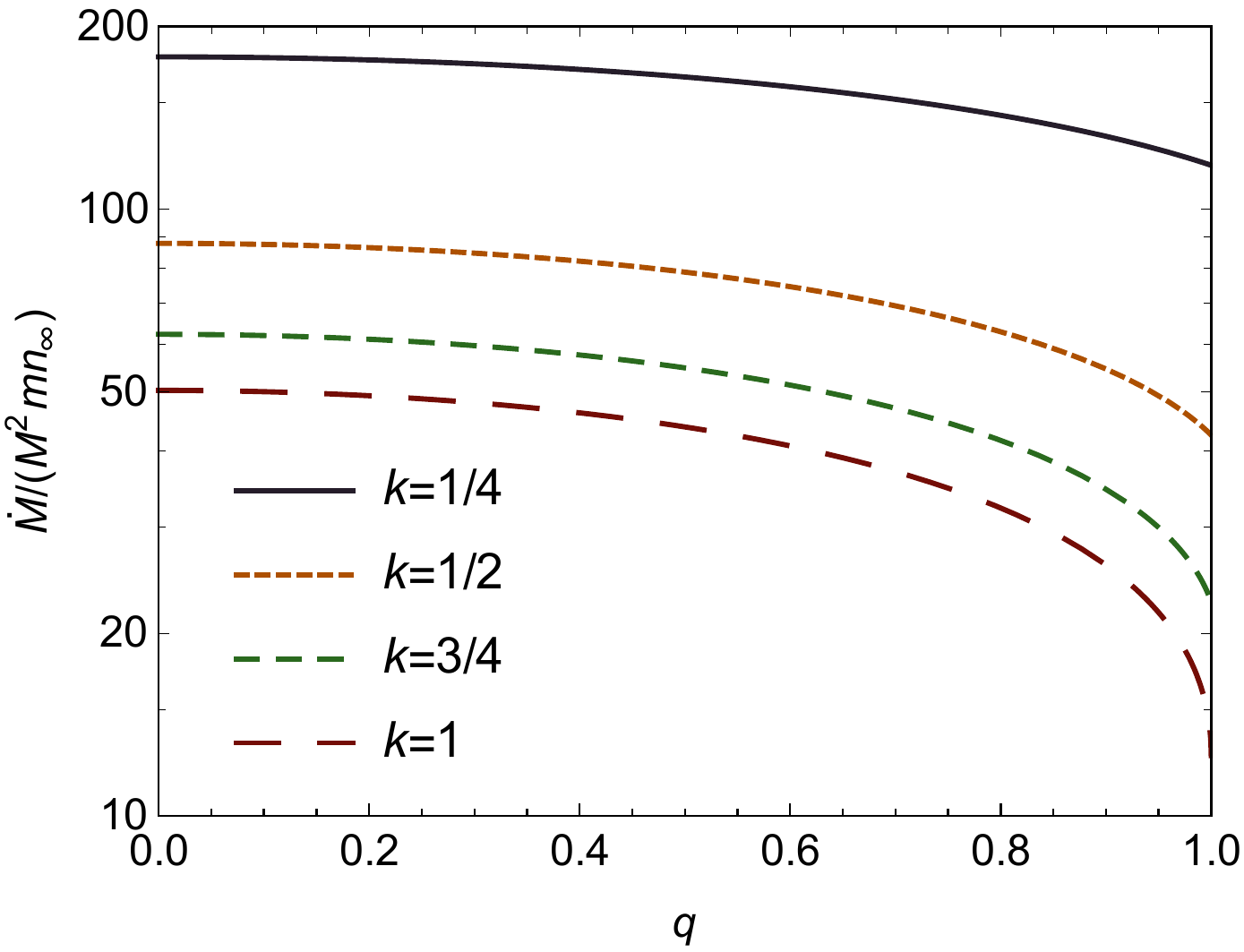}
    \caption{Mass accretion rate $\dot M/(M^2 m n_\infty)$ vs.\ the charge parameter $q$ for the perfect-fluid model with the linear equation of state $p = k \rho$.}
    \label{fig:mdot3}
\end{figure}

\begin{figure}
    \centering
    \includegraphics[width=0.45\textwidth]{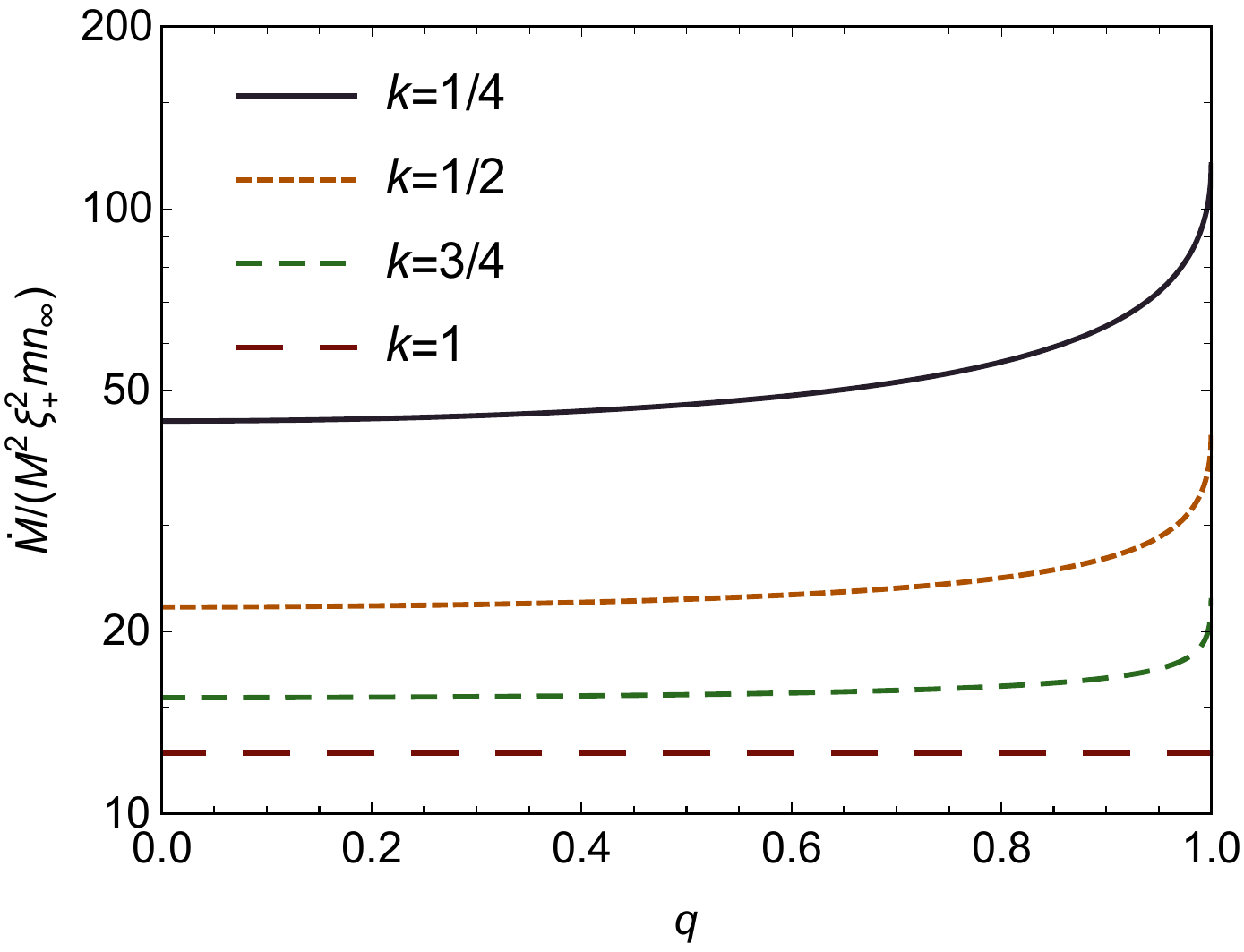}
    \caption{Mass accretion rate $\dot M/(M^2 \xi_+^2 m n_\infty)$ vs.\ the charge parameter $q$ for the perfect-fluid model with the linear equation of state $p = k \rho$.}
    \label{fig:mdot4}
\end{figure}

Stationary, spherically symmetric accretion of perfect fluids in the Reissner-Nordstr\"{o}m spacetime was studied e.g.\ in \cite{babichev,ficek}. It is a simple model that can serve as a reference for the results of this paper. In this appendix we give a few formulas that can be obtained for the linear equation of state of the form $p = k \rho$, where $\rho$ is the energy density, $p$ is the  pressure, and $0 < k \le 1$ is a constant (the square of the local speed of sound).

We assume standard conditions of the relativistic Bondi-type accretion: The flow is stationary and spherically symmetric. We assume the energy-momentum tensor of the perfect fluid of the form (\ref{tmunu_perf_fluid}).
The Reissner-Nordstr\"{o}m metric is given by Eq.\ (\ref{rn_eddington_finkelstein}). The four-velocity of the fluid satisfies $u^\theta = u^\varphi = 0$. Due to the symmetry assumptions $p$, $\rho$, $u^t$, $u^r$ can only depend on the radius $r$. 

The conservation equations $\nabla_\mu(nu^\mu) = 0$, $\nabla_\mu T\indices{^\mu_\nu} = 0$ yield
\begin{eqnarray}
\label{perfect_fluid_accretion_rate}
r^2 n u^r & = & - \frac{\dot M}{4 \pi m}, \\
h \sqrt{1 - \frac{2M}{r} + \frac{Q^2}{r^2} + (u^r)^2} & = & h_\infty,
\end{eqnarray}
where $\dot M$ and $h_\infty$ are constants. The enthalpy per particle $h$ is defined as $h = (p + \rho)/n$, and $h_\infty$ is its asymptotic value.

For the linear equation of state $p = k \rho$ we get
\[ n = n_\infty \left( \frac{\rho}{\rho_\infty} \right)^\frac{1}{1+k}, \]
\[ h = \frac{(1+k) \rho_\infty}{n_\infty^{1+k}} n^k, \]
where $n_\infty$ and $\rho_\infty$ denote the asymptotic values of $n$ and $\rho$, respectively.

This yields
\begin{equation}
\label{perfect_fluid_accretion_n}
n^k \sqrt{1 - \frac{2M}{r} + \frac{Q^2}{r^2} + (u^r)^2} = n^k_\infty.
\end{equation}

Combining Eqs.\ (\ref{perfect_fluid_accretion_n}) and (\ref{perfect_fluid_accretion_rate}) we get the following, general expression for the mass accretion rate:
\begin{equation}
\label{perfect_fluid_accretion_rate_general}
\dot M = - \frac{4 \pi m r^2 u^r n_\infty}{\left[ 1 - \frac{2 M}{r} + \frac{Q^2}{r^2} + (u^r)^2 \right]^\frac{1}{2k}}.
\end{equation}

In the following, we restrict ourselves to critical solutions, i.e., solutions passing through a (saddle-type) critical point. These are the only solutions joining smoothly the black-hole horizon with the infinity.  Characteristics of the critical solutions can be obtained from the following relations:
\begin{equation}
\label{critical_point}
(u^r_\ast)^2 = \frac{M}{2r_\ast} - \frac{Q^2}{2 r_\ast^2} = k \left( 1 - \frac{3 M}{2 r_\ast} + \frac{Q^2}{2 r_\ast^2} \right),
\end{equation}
in which the quantities referring to the critical point are denoted with the asterisk.

Solving Eqs.\ (\ref{critical_point}) with respect to $r_\ast$ and $u^r_\ast$, and inserting the results into Eq.\ (\ref{perfect_fluid_accretion_rate_general}), we get
\[ \dot M = \pi m M^2 k^{-\frac{3}{2}} 2^{-\frac{1 + 3k}{2k}} Y^\frac{k+1}{k} Z^\frac{k-1}{2k} M^2 n_\infty, \]
where
\begin{eqnarray*}
Y & = & 1 + 3k \\
&& + \sqrt{1 + k (6 - 8q^2) + k^2 (9 - 8q^2)}, \\
Z & = & 1 + k (3 - 4 q^2) \\
&& + \sqrt{1 + k (6 - 8q^2) + k^2 (9 - 8q^2)}. \\
\end{eqnarray*}
In particular, for $k = 1$, one obtains
\[ \dot M = 4 \pi m M^2 \left( 1 + \sqrt{1 - q^2} \right)^2 n_\infty = 4 \pi m M^2 \xi_+^2 n_\infty. \]

In Figs.\ \ref{fig:mdot3} and \ref{fig:mdot4} we plot the ratios $\dot M/(M^2 m n_\infty)$ and $\dot M/(M^2 \xi_+^2 m n_\infty)$, respectively. These figures can be compared with Figs.\ \ref{fig:mdot1} and \ref{fig:mdot2} obtained for the Vlasov gas. Similarly to the Vlasov gas, the ratio $\dot M/(M^2 m n_\infty)$ decreases with $q$; the ratio $\dot M/(M^2 \xi_+^2 m n_\infty)$ (normalization by the area of the horizon) increases with $q$ for $0 < k < 1$. In the limiting case of ultra-stiff fluids ($k = 1$), $\dot M/(M^2 \xi_+^2 m n_\infty) = \mathrm{const} = 4 \pi$.

Analytic solutions of the form $n = n(r)$, $u^r = u^r(r)$, etc, can be obtained for selected values of $k = 1/4, 1/2, 1/3, 1, \dots$, however in most cases the corresponding formulas are lengthy. The cases with $k = 1/2$ and $k = 1$ are exceptional. For $k = 1/2$ we get
\[ u^r = - \frac{1}{2 \kappa} \left[ r^2 \pm \sqrt{r^4 - 4 \kappa^2 \left( 1 - \frac{2M}{r} + \frac{Q^2}{r^2} \right)} \right], \]
where $\kappa = \dot M/(4 \pi m n_\infty)$. The result for $k = 1$ reads
\[ (u^r)^2 = \frac{\kappa^2 \left( 1 - \frac{2M}{r} + \frac{Q^2}{r^2} \right)}{r^4 - \kappa^2}. \]
For the critical flow with $k = 1$ (the only solution regular at the black hole horizon) we get
\[ \left( \frac{n}{n_\infty} \right)^2 = \frac{(\xi + \xi_+)(\xi^2 + \xi_+^2)}{\xi^2 (\xi - \xi_-)}. \]
Consequently, the ratio of $n/n_\infty$ at the black hole horizon (the compression factor) is given by
\[ \frac{n}{n_\infty} = \sqrt{2} \left(1 + \frac{1}{\sqrt{1 - q^2}} \right)^\frac{1}{2}, \]
and it grows with $q$.

\end{document}